\pdfoutput=1
\documentclass[aps,prb,superscriptaddress,nofootinbib,twocolumn,longbibliography,10pt]{revtex4-2}
\usepackage{amsmath,amssymb,amsfonts,graphicx,bm,color,floatrow}
\usepackage[colorlinks=true,allcolors=blue]{hyperref}
\usepackage[caption=false]{subfig}

\newcommand{\beq}{\begin{eqnarray}}
\newcommand{\eeq}{\end{eqnarray}}
\newcommand{\non}{\nonumber\\}
\DeclareMathOperator{\U}{U}

\newcommand{\p}{\partial}
\renewcommand{\i}{\mathrm{i}}
\renewcommand{\d}{\mathop{}\!\mathrm{d}}

\newcommand{\calE}{\mathcal{E}}

\newcommand{\calP}{\mathcal{P}}
\newcommand{\calQ}{\mathcal{Q}}

\newcommand{\bX}{\mathbf{X}}

\newcommand{\be}{\mathbf{e}}
\renewcommand{\bm}{\mathbf{m}}
\newcommand{\bn}{\mathbf{n}}

\newcommand{\figsfolder}{figs/}

\begin{document}
\title{From Nested Soliton Crystals to Skyrmion Crystals in Chiral Magnets}
\date{July, 2025}
\author{Sven Bjarke Gudnason}
\affiliation{Institute of Contemporary Mathematics, School of
  Mathematics and Statistics, Henan University, Kaifeng, Henan 475004,
  P.~R.~China}
\affiliation{Department of Physics, Chemistry and Pharmacy,
  University of Southern Denmark, Campusvej 55, 5230 Odense M,
  Denmark}
\author{Muneto Nitta}
\affiliation{Department of Physics, Keio University, 4-1-1 Hiyoshi,
  Yokohama, Kanagawa 223-8521, Japan}
\affiliation{Research and Education Center for Natural Sciences, Keio
  University, 4-1-1 Hiyoshi, Yokohama, Kanagawa 223-8521, Japan}
\affiliation{International Institute for Sustainability with Knotted
  Chiral Meta Matter(SKCM2 ), Hiroshima University, 1-3-1 Kagamiyama,
  Higashi-Hiroshima, Hiroshima 739-851, Japan}
\begin{abstract}
We perform an exhaustive study of the parameter space of the chiral
magnetic model with a Dzyaloshinskii-Moriya interaction, an in-plane
anisotropy term and an out-of-plane Zeeman magnetic field.
We map out the phase diagram for the model and find a chiral soliton
lattice hosted on a chiral soliton lattice as the ground state in a particular
corner of the phase diagram.
There is a smooth crossover, as opposed to a phase transition, to the
skyrmion crystal lattice phase.
Finally, we also discover excited composite soliton states such as an
anti-skyrmion coexisting with a bimeron or an anti-bimeron.
\end{abstract} 
\maketitle

\section{Introduction}

Chiral magnets are fascinating physical systems that host a wide array
of solitons -- extended pseudo-particles -- in various different
phases, depending on the magnitude of the magnetic field and the
anisotropy coefficient of the material in question.
The most famous soliton in chiral magnets is perhaps the skyrmion
\cite{1989JETP...68..101B,1995JETPL..62..247B,Nagaosa2013}, which is a
2-dimensional texture with topological charge characterized by the
second homotopy group, since its target space is a 2-sphere.
This can be understood from the fact that ``hard'' magnets have an
approximately constant magnetic field, but the magnetization vector
can be twisted around in three directions: this forms a 2-sphere.
This skyrmion is a lower-dimensional analogue of the original skyrmion
that originated in chiral perturbation theory in particle and nuclear
physics, which not only is a 3-dimensional texture, but also has a
3-dimensional target space \cite{Skyrme:1961vq,Skyrme:1962vh}.
Skyrmions do not exist with only a kinetic term due to a virial law
also known as Derrick's theorem \cite{Derrick:1964ww}.
In three dimensions the skyrmions' stability requires the Skyrme term
(or a similar higher-derivative term), whereas in two dimensions the
skyrmions require the Dzyaloshinskii-Moriya interaction (DMI)
\cite{DZYALOSHINSKY1958241,Moriya:1960zz},
which in distinction with the 3-dimensional skyrmions breaks parity
symmetry.
Magnetic skyrmions have been realized in the laboratory in MnSi
\cite{doi:10.1126/science.1166767} as well as in
$\text{Fe}_x\text{Co}_{1-x}\text{Si}$ \cite{Yu2010} 
as skyrmion crystals (SkX)
\cite{doi:10.1126/science.1166767,Yu2010,doi:10.1038/nphys2045,Rossler:2006,Han:2010by,Lin:2014ada,Ross:2020hsw}
and are popular
candidates for low-energy consumption data-storage devices
\cite{Fert2013}. 

Their technological breakthrough has, however, been hampered by the
so-called skyrmion-Hall effect 
\cite{zang2011dynamics,Chen:2017,Jiang:2017,litzius2017skyrmion} 
that bends the skyrmion's trajectories
thus making it difficult to control the precise positions that should
correspond to bits in a data-storage device.
Two possible ways to overcome this obstacle are anti-ferromagnets and
domain-wall(DW) skyrmions.
The former has a relativistic kinetic term for the pseudo-particles
(skyrmions) due to cancellation of the leading order lattice expansion
of the Heisenberg interaction, which hence changes the dynamics from
Landau-Lifshitz-Gilbert dynamics to relativistic dynamics (albeit with
a lower ``speed of light'').
The latter possibility is what we will consider here: the
DW-skyrmion.

The DW-skyrmion is a composite soliton where  
a skyrmion is absorbed into 
a DW~\cite{Kim:2017lsi,PhysRevB.99.184412,PhysRevB.102.094402,Kuchkin:2020bkg,Ross:2022vsa,Lee:2022rxi,Amari:2023gqv,Amari:2023bmx,Gudnason:2024shv,Leask:2024dlo,Lee:2024lge,Xu2024,Gudnason:2025gqs,
Chen_2025,
PhysRevB.109.014404,Nie_2025,PhysRevB.111.024402,b2s4-d69w,chen2025currentinduceddynamicsblochdomainwall} 
and this has been experimentally observed 
\cite{Nagase2021,10.1063/5.0056100,Yang2021,Bruning2025,Yang2026}.
The clear advantage of the magnetic DW-skyrmion is that it, just like a train
on a metal track, is bound to move only along the DW -- this way the
bent trajectories are avoided and the skyrmion-Hall effect is no
longer a problem.
DW-skyrmions originate from quantum field theory where 
a 2- or 3-dimensional skyrmion is confined in a co-dimension 1 DW in 
two \cite{Nitta:2012xq,Kobayashi:2013ju,Jennings:2013aea} or
three 
\cite{Nitta:2012wi,Nitta:2012rq,Gudnason:2014nba,Gudnason:2014hsa,Eto:2015uqa,Nitta:2022ahj} 
spatial dimensions, respectively. 
While the former is a prototype of magnetic DW-skyrmions,
the latter has recently been found to support a new phase of quantum
chromodynamics in the presence of a strong magnetic field
\cite{Eto:2025fkt,Eto:2023wul,Amari:2024fbo,Copinger:2025rpo,Copinger:2026col} or a rapid rotation \cite{Eto:2023tuu}.\footnote{
The term ``DW-skyrmion'' was coined in Ref.~\cite{Eto:2005cc} as a Yang-Mills instanton in a DW. With the terminology used in this paper, it
should have been called a ``DW-instanton''.}

The DW-skyrmions with a out-of-plane anisotropy term bend into a cusp
and which side of the DW the cusp appears on depends on the DW versus
anti-DW hosting a skyrmion versus an anti-skyrmion \cite{Kim:2017lsi,PhysRevB.99.184412,PhysRevB.102.094402,Kuchkin:2020bkg,Ross:2022vsa,Lee:2022rxi,Amari:2023gqv,Amari:2023bmx}.
In this paper, however, we study the DW-skyrmion in presence of an
out-of-plane Zeeman magnetic field and an \emph{in-plane} anisotropy
term, for previous work on this model see \cite{Amari:2024jxx}.
In this case, the DW-skyrmion does not bend to either side, but
appears symmetrically on the DW.

The DW only exists for large enough ratio of the anisotropy
coefficient ($K$) to the Zeeman magnetic field ($B$); their
dimensionless parameters, where the exchange energy and DMI
coefficient have been rescaled into these parameters, are defined in
the next section.
In this paper, we perform an exhaustive study of the phase diagram in 
the $K$-$B$ parameter space, which uncovers a new excited state, which
is an anti-bimeron that lives on the DW as well as another new excited
state that is the anti-skyrmion living on the anti-DW.
For the definitions and termology of the soliton versus its
anti-soliton, see the next section.
The bimeron in our terminology is different from the skyrmion, since
it is a DW-skyrmion that in some part of the parameter space is
deconfined into two halves of the topological unit charge that the
skyrmion usually carries.
It is also bigger than the skyrmion (or anti-skyrmion).
One may also interpret the bimeron as two confined ends of DWs.
This becomes apparent in some parts of the parameter space, where the
gap between two DWs increases and the bimeron becomes almost
unconfined.

A main result of our study is the discovery of the DW-anti-DW ground
state (also known as the chiral soliton lattice (CSL) 
\cite{PhysRevLett.108.107202,KISHINE20151,PhysRevB.97.184303,Ross:2020orc}) with a
kink lattice ground state hosted on the DW (and not on the anti-DW):
We call this composite ground state a CSL on a CSL.
The region in parameter space where the CSL on a CSL exists is,
however, so close to the skyrmion crystal lattice phase (SkX)
\cite{doi:10.1126/science.1166767,Yu2010,doi:10.1038/nphys2045,Rossler:2006,Han:2010by,Lin:2014ada,Ross:2020hsw} 
that
visually it is difficult to see the difference between the CSL on a
CSL and the hexagonal lattice of skyrmions in the skyrmion crystal
lattice phase.
It is dictated by the mathematical conditions and not by empirically
observing how the solutions look like.
Indeed, we find that there is no phase transition between the CSL on a
CSL phase to the skyrmion crystal lattice phase: It is a smooth
crossover.

Perhaps the most interesting result is the discovery of the
anti-skyrmion excited states living on the anti-DW that can be
coexisting with the bimeron on the DW, with the latter either as the
ground state on the DW or an excited state (in the ferromagnetic phase).
The inter-soliton interactions give rise to aligned systems in the
direction perpendicular to the DW, which is in contradistinction to
the hexagonal patterns that emerge for the pure skyrmion phase or for
the pure bimeron phase.
We also find for the first time, to the best of our knowledge, an even
higher-energy excited state which consists of the anti-skyrmion on the
anti-DW coexisting with the anti-bimeron living on the DW.
Again we find an alignment giving rise to rectangular lattices as
opposed to hexagonal lattice.

This paper is organized as follows.
In the next section, we introduce our notation for the model and
convert it to dimensionless units, defining the two parameters that
span the parameter space, which we study in detail in
Sec.~\ref{sec:results} with the new states and the phase diagram as
the main results.
We conclude in Sec.~\ref{sec:conclusion} with a discussion and
outlook.

\section{Chiral magnetic model with in-plane anisotropy and out-of-plane Zeeman}

The energy functional for the chiral magnetic model with in-plane
anisotropy and out-of-plane Zeeman potential reads
\begin{align}
  E &= \int\widetilde{\calE}\d^2\tilde{x},\\
  \widetilde{\calE} &= A\,\p_i\bn\cdot\p_i\bn
  + D\bn\cdot\nabla\times\bn
  + \mathsf{K}(1-n_1^2)
  + \mathsf{B}(1-n_3),\nonumber
\end{align}
where $\widetilde{\calE}$ and $\tilde{x}$ are the dimensionful energy
density and coordinate in the plane, $A$ is the exchange energy,
$\p_i$, $i=1,2$ is the partial derivative in the plane,
$\bn=\bm/M_{\rm sat}$ is the normalized magnetization field,
$\bn=(n_1,n_2,n_3)$, $\bn\cdot\bn=1$, $D$ is
the DMI coefficient, $\nabla=(\p_1,\p_2,0)$, $\mathsf{K}$ is the
in-plane anisotropy coefficient and finally $\mathsf{B}$ is the
magnetic field inducing the out-of-plane Zeeman term.
Rescaling the lengths and measuring the energy in units of $2A$, we
can write the model as
\begin{align}
  E &= 2A\int\calE\d^2x,\non
  \calE &= \frac12\p_i\bn\cdot\p_i\bn
  + \bn\cdot\nabla\times\bn
  + V,\\
  V &=
  K(1-n_1^2)
  + B(1-n_3),\nonumber
\end{align}
with only two dimensionless parameters: $K$ and $B$ that measure the
effective in-plane anisotropy and effective out-of-plane Zeeman.
They are related to the physical constants as
\beq
K = \frac{2A\mathsf{K}}{D^2}, \qquad
B = \frac{2A\mathsf{B}}{D^2},
\eeq
and both $\calE$ and $x$ are dimensionless.
Without loss of generality, we can assume that $B\geq0$
\cite{Barton-Singer:2024clx} and for now we will leave
$K\neq 0$ as a nonvanishing real parameter, although we will mostly focus
on the case where $K>0$ which is necessary (but not sufficient) for
the existence of DWs.

The static equations of motion (EOM) are
\begin{equation}
{\rm eom}_{\bn}=
\calP\left(\nabla^2\bn
- 2\,\nabla\times\bn
+ 2Kn_1\be_1
+ B\,\be_3\right) = 0,
\label{eq:eom}
\end{equation}
where $\calP(\bX)=\bX-(\bn\cdot\bX)\bn$ is a projection operator to the
target space 2-sphere.

\subsection{Vacua}

The vacuum equation is obtained from the EOM \eqref{eq:eom} by setting
the derivatives to zero and can be written as
\begin{align}
  \cos\theta\left(2K\sin\theta - B\cos\phi\right) = 0,\\
  B \sin\theta\sin\phi = 0,
\end{align}
where we have assumed the Ansatz for $\bn$:
\beq
\bn = \begin{pmatrix}
  \cos\theta\\
  -\sin\theta\sin\phi\\
  \sin\theta\cos\phi
  \end{pmatrix},
\label{eq:theta_phi_ansatz}
\eeq
where $\theta\in[0,\pi]$ and $\phi\in[0,2\pi)$ are the standard
spherical coordinates on the 2-sphere target space.
If $B=0$, the two vacua are:
\beq
\theta = 0 \quad\textrm{or}\quad \theta=\pi, \qquad
\phi\in[0,2\pi), \qquad
V = 0,
\eeq
for $K>0$ and correspond to DW vacua with a $\U(1)$ modulus,
whereas for $K<0$:
\begin{equation}
\theta=\frac{\pi}{2}\quad\textrm{or}\quad \theta=\frac{3\pi}{2}, \qquad
\phi\in[0,2\pi), \qquad
V = K < 0,
\end{equation}
and $V$ is the potential energy density in the given vacuum.

We will now turn to the case of $B>0$ ($B$ is non-negative) where we
without loss of generality can set $\phi:=0$.
If $K<0$ is negative, the vacuum is now
\beq
\theta=\frac{\pi}{2}, \qquad
V = K < 0,
\label{eq:single_vac}
\eeq
which only has one vacuum and therefore no DW exists.
Since we are interested in DWs we will take $K>0$ from hereon, in
which case there are two possibilities:
If $B<2K$ two shifted DW vacua have the lowest energy
\begin{align}
\theta &= \arcsin\left(\frac{B}{2K}\right), \qquad&
V &= B\left(1 - \frac{B}{4K}\right),\label{eq:vac1}\\
\theta &= \pi - \arcsin\left(\frac{B}{2K}\right), \qquad&
V &= B\left(1 - \frac{B}{4K}\right),\label{eq:vac2}
\end{align}
whereas if the (effective) magnetic field surpasses the critical value
set by the (effective) anisotropy, $B\geq 2K$, only the vacuum
\eqref{eq:single_vac} exists which does not support DWs.
Indeed, the two above vacua coalesce in the limit $B\to 2K$.

\subsection{DWs}

The existence of DWs requires a disjoint set of two or more degenerate
vacua in order for $\pi_0(\mathcal{V}_{\rm vac})$ not to be trivial.
In the case of $K>\frac12B\geq0$, two disjoint degenerate vacua were
found in the previous subsection.
In the previous subsection, we also found that we can set $\phi:=0$
in the Ansatz \eqref{eq:theta_phi_ansatz} without loss of generality
in the DW sector.
Promoting $\theta$ from a coordinate on the target space to a field
variable $\theta=\theta(y)$ ($x=x^1$ and $y=x^2$), we can write the
DW energy density as 
\beq
\calE' = \frac12(\theta')^2 + \theta'
+ \frac{\left(B - 2K\sin\theta\right)^2}{4K},
\eeq
where we have subtracted the negative DW energy density off of $\calE$
as $\calE'=\calE - B + \frac{B^2}{4K}$ and we denote the
$y$-derivative as $\theta'=\frac{\d\theta}{\d y}$.
The shift of the energy functional by a constant does not affect the
vacuum conditions or EOM.
The EOM for the DW with the Ansatz is
\beq
\theta'' + (B - 2K\sin\theta)\cos\theta = 0,
\label{eq:DW_eom}
\eeq
which for $B=0$ is the standard sine-Gordon equation, whereas for
$B>0$ is the so-called double sine-Gordon equation
\cite{Burt:1977ii,PhysRevB.27.474}.
Although the EOM is unaffected by the presence of the DMI, the energy
density induces a potential on the DW depending on its ``orientation''
(i.e.~value of $\phi$ in Eq.~\eqref{eq:theta_phi_ansatz}).

There exist four solutions to the EOM \eqref{eq:DW_eom}.
We denote the first two as the short DWs \cite{Amari:2024jxx}
\beq
\theta_S^\pm = \pm2\arctan\left(\alpha\tanh\left[\beta(y-y_0)\right]\right) + \frac{\pi}{2},
\label{eq:thetaSpm}
\eeq
whereas the long DWs are given by \cite{Amari:2024jxx}
\beq
\theta_L^\pm = \pm2\arctan\left(\frac{1}{\alpha}\tanh\left[\beta(y-y_0)\right]\right) - \frac{\pi}{2},
\label{eq:thetaLpm}
\eeq
where we have defined
\beq
\alpha = \sqrt{\frac{1-\gamma}{1+\gamma}}, \quad
\beta = \sqrt{\frac{K(1-\gamma^2)}{2}}, \quad
\gamma = \frac{B}{2K}.
\label{eq:alpha_beta_gamma}
\eeq
The DW solutions versus anti-DW solutions are
\beq
\begin{cases}
  \textrm{DW} : & \theta_S^+,\theta_L^-,\\
  \textrm{anti-DW} : & \theta_S^-, \theta_L^+.
\end{cases}
\eeq
The energies of the DWs are given by \cite{Amari:2024jxx}
\begin{align}
  E'{}_S^\pm &= 4L_x\left[\beta + \left(\pm1 - \frac{B}{\sqrt{2K}}\right)\arctan(\alpha)\right],\label{eq:ESpm}\\
  E'{}_L^\pm &= 4L_x\left[\beta + \left(\pm1 + \frac{B}{\sqrt{2K}}\right)\arctan\left(\frac{1}{\alpha}\right)\right],\label{eq:ELpm}
\end{align}
where $L_x$ is the length of the DW (in the $x$-direction).

It is instructive to consider the limit $B\to0$ ($K>0$) in the
analytic solutions.
In this limit $\alpha\to1$, $\beta\to\sqrt{\frac{K}{2}}$ and
$\gamma\to0$.
Using the identity
\beq
\arctan(u)-\arctan(v)=\arctan\left(\frac{u-v}{1+uv}\right),
\eeq
valid for $-1<uv$, and expressing $\tanh\big(\sqrt{K/2}(y-y_0)\big)$ as
$\frac{u-v}{1+uv}$ with $u=e^{\sqrt{2K}(y-y_0)}$ and $v=1$, we see
that the short walls are indeed just he renowned sine-Gordon solution
\beq
\theta_S^{\pm}\to\pm2\arctan\left(\exp\left[\sqrt{2K}(y-y_0)\right]\right),
\eeq
and the long walls are shifted downwards by $\pi$.

A degenerate point exists where $E'{}_L^-=E'{}_S^+$ which can be
analytically established and is at:
\beq
K = \frac12B^2, \qquad B>1.
\label{eq:degenerate_point}
\eeq
This can be seen from the vanishing of both parentheses in front of
$\arctan$ in Eqs.~\eqref{eq:ESpm} for the $+$ sign and \eqref{eq:ELpm}
for the $-$ sign.
A little analysis reveals that for $K>\frac12B^2$, then
$E'{}_L^-<E'{}_S^+$ whereas for $K<\frac12B^2$ we have
$E'{}_L^->E'{}_S^+$ \cite{Amari:2024jxx}.

The DW exists for $K>\frac12B$, so a small $K$ is not applicable
for fixed $B$; on the other hand, it may prove useful for expand the
energies in small $B/K$:
\begin{align}
  E'{}_S^\pm &= L_x\left[2\sqrt{2K} - \frac{\pi B}{\sqrt{2K}} \pm \left(\pi-\frac{B}{K}\right)\right] + \mathcal{O}\left(\frac{B^2}{K^{\frac32}}\right),\non
  E'{}_L^\pm &= L_x\left[2\sqrt{2K} + \frac{\pi B}{\sqrt{2K}} \pm \left(\pi+\frac{B}{K}\right)\right] + \mathcal{O}\left(\frac{B^2}{K^{\frac32}}\right),\label{eq:EDW_L_expansion}
\end{align}
so for very large $K/B$ the short and long DWs become degenerate in
energy with the $\pm$ walls each being separated by a constant
relative shift of $2\pi L_x$.
Consistently, we can also see from the expansion
\eqref{eq:EDW_L_expansion} that $K=B^2/2$
(Eq.~\eqref{eq:degenerate_point}) indeed gives rise to a degenerate  
point in energy between $E'{}_S^+$ and $E'{}_L^-$.

\subsection{Bulk CSL}

We can see from Eq.~\eqref{eq:EDW_L_expansion} that the DW energies
are positive for large values of $K$, whereas for the CSL we need the
DW energies to be negative (recall that the homogeneous phase's energy
is subtracted off in $E'$), hence we need to consider small $K$ and small
$B$ (since $K>B/2$ is necessary for the existence of DWs).

It turns out that the CSL exists in the region of parameter space
where $K>\frac12B^2$ so the DW energy hierarchy is given by
$E'{}_L^-<E'{}_S^+$.
Since the energy of the anti-DW $E'{}_S^-$ turns out to be negative in
a much larger region of parameter space than the DW $E'{}_L^-$, the
latter gives rise to the strong condition for the CSL in the bulk.
This strong condition for CSL (i.e.~$E'{}_L^-<0$) is analytic but not
explicit in $K$ and $B$:
\beq
\beta - \left(1 - \frac{B}{\sqrt{2K}}\right)
\arctan\left(\frac{1}{\alpha}\right) < 0,
\label{eq:CSL_strong_cond}
\eeq
where $\alpha$ and $\beta$ are given in
Eq.~\eqref{eq:alpha_beta_gamma}.

We will also consider a weaker condition for bulk CSL, which is that
we allow the $\theta_L^-$ DW to have a positive energy as long as the
sum of the energies of the DW, $\theta_L^-$, and the anti-DW,
$\theta_S^-$, is negative;
i.e.~$E'{}_L^-+E'{}_S^-<0$, which can be written as:
\beq
2\beta - \frac{\sqrt{2}B}{\sqrt{K}}\arctan(\alpha)
- \frac\pi2\left(1-\frac{B}{\sqrt{2K}}\right)
< 0,
\label{eq:CSL_weak_cond}
\eeq
where $\alpha$ and $\beta$ are given in
Eq.~\eqref{eq:alpha_beta_gamma}. 

\subsection{Topological charge}

The DWs possess a topological charge only when fixing the boundary
conditions such that the vacuum at $y\to-\infty$ is different from
that of $y\to\infty$ for every $x$.
Hence the topology is imposed by the boundary conditions or in other
words, the DW is a codimension-one soliton in a two-dimensional plane. 

There is a topological charge of the map
$\mathbb{R}^2\to\bn$, when the boundary condition
$\lim_{r\to\infty}\bn=\be_3$ (with $x+\i y=r e^{\i\varphi}$), which 
point-compactifies the plane to a 2-sphere and hence the topological
charge or skyrmion number is $\pi_2(S^2)=\mathbb{Z}\ni Q$, with
\beq
Q = -\frac{1}{4\pi}\int \bn\cdot\p_1\bn\times\p_2\bn\;\d^2x.
\eeq
Inserting the Ansatz \eqref{eq:theta_phi_ansatz} and assuming that
$\theta=\theta(y)$ and $\phi=\phi(x)$, we get the topological charge
\beq
Q = \frac{1}{4\pi}\left[-\cos\theta\right]_{\theta(y_{\rm min})}^{\theta(y_{\rm max})}
\left[\phi\right]_{\phi(x_{\rm min})}^{\phi(x_{\rm max})}.
\eeq
With the topological charge formula in hand, let us next consider
initial conditions for numerical computations.

\subsection{Initial configurations}\label{sec:initial_conf}

The vacua \eqref{eq:vac1}-\eqref{eq:vac2} with $\phi=0+2\pi n$
allow for a kink on the DW, which we want to call a CSL on a CSL
(although the conditions for this to be true will only be uncovered
later on).
The kink and anti-kink have the boundary conditions
\beq
\lim_{x\to-\infty}\phi=0, \qquad
\lim_{x\to\infty}\phi=\pm 2\pi,
\eeq
with the upper sign for the kink and the lower sign for the anti-kink.
For concreteness we will use the initial condition in the numerical
computations given by the Ansatz \eqref{eq:theta_phi_ansatz} with
$\theta$ being $\theta_{L,S}^{\pm}$ of
Eqs.~\eqref{eq:thetaSpm}-\eqref{eq:thetaLpm} and the (anti-)kink given
by
\beq
\phi = 4 \arctan\left(e^{\pm m x}\right),
\label{eq:phi_ansatz}
\eeq
with a suitably chosen kink mass $m$.

We might think that putting a kink on a DW will correspond to a
skyrmion of charge $Q=1$, but this is slightly modified due to the
shifted vacua \eqref{eq:vac1}-\eqref{eq:vac2}.
Indeed, evaluating the topological charge for the DWs and anti-DWs
(which just requires their asymptotic values, viz.~vacua), we obtain
\beq
Q = \pm\pm'Q_{\rm sh}, \qquad
Q_{\rm sh} = \sqrt{1-\frac{B^2}{4K^2}},
\label{eq:Qsh}
\eeq
where the $\pm$ is $+$ for the DW and $-$ for the anti-DW
and the $\pm'$ is $+$ for the kink and $-$ for the anti-kink.
For $B\to 0$, the topological charge or skyrmion number is indeed an
integer, but for the shifted vacua \eqref{eq:vac1}-\eqref{eq:vac2} the
topological charge decreases due to the fact that the 2-sphere is not
completely covered by the skyrmion map -- the shifted vacua leave
small discs uncovered (or doubly covered, giving rise to the same effect) at the north and south poles of the 2-sphere.

\begin{table}[!htp]
  \begin{center}
    \begin{tabular}{r||cc}
      & kink & anti-kink\\
      \hline\hline
      DW & skyrmion & anti-skyrmion\\
      anti-DW & anti-skyrmion & skyrmion
    \end{tabular}
    \caption{DW and kink charges and their relation to the topological
      charge, which is the skyrmion number. Due to the shifted vacua
      \eqref{eq:vac1}-\eqref{eq:vac2} the skyrmion's topological charge is
      $Q=Q_{\rm sh}$, whereas that of the anti-skyrmion
      is $Q=-Q_{\rm sh}$ with $Q_{\rm sh}$ defined in
      Eq.~\eqref{eq:Qsh}. 
      }
    \label{tab:Q}
  \end{center}
\end{table}
The product of the signs means skyrmions and anti-skyrmions are
related to the DW and kinks as shown in Tab.~\ref{tab:Q}.
Although the topological charge is the same for the long and the short
DWs, the energies are not the same -- as can be seen from
Eqs.~\eqref{eq:ESpm}-\eqref{eq:ELpm}.
Indeed, not every combination of the 2 DWs and 2 anti-DWs with the
kink and the anti-kink (in principle 8 different combinations) gives
rise to a stable soliton solution with a(n anti-)skyrmion, as we will
see next.

\subsection{Low-energy effective theory in the moduli space approximation}

The low-energy effective theory in the moduli (or Manton)
approximation 
\cite{Manton:1981mp,Eto:2006uw}
has been derived in Ref.~\cite{Amari:2024jxx} using the 
approximation of vanishing Zeeman field (i.e.~the limit of $B\to0$).
Indeed, the moduli space approximation for this system is very subtle
due to the fact of the complicated and shifted vacuum structure.
The issue is this.
Normally, one would insert the solution of the heavy soliton, say
$\theta_S^+$ into the Hamiltonian and integrate over the codimension
of the DW to obtain and effective action for the quasi-modulus $\phi$
an the modulus $y_0$.
However, the coefficient of the kinetic term for the quasi-modulus
$\phi$ diverges due to the incompatibility of the Ansatz and the vacua
for $B\neq 0$, that is, $\int\sin^2(\theta)\d y\to\infty$ for $B\neq 0$,
which is the coefficient of $\frac12(\p_x\phi)^2$.
This issue was dealt with in Ref.~\cite{Amari:2024jxx} by assuming
that $B\to0$ for the low-energy effective Hamiltonian.

It is worthwhile to mention a difference in the moduli space approximation of Ref.~\cite{Amari:2024jxx} with respect to our Ansatz \eqref{eq:phi_ansatz}: The moduli space approximation generically gives rise to the double-sine Gordon solution, whereas our Ansatz in Eq.~\eqref{eq:phi_ansatz} is simply the sine Gordon solution (that has different vacua compared with the latter).
This is due to the moduli space approximation in Ref.~\cite{Amari:2024jxx} being carried out for the more general DMI with a mixing parameter $\tilde\vartheta$ that mixes the Bloch- (corresponding to $\tilde\vartheta=0$) and N\'eel-type (corresponding to $\tilde\vartheta=\pi/2$) DMI terms.
When the DMI-term is not of Bloch-type (i.e.~$\tilde\vartheta=0$) then the appropriate Ansatz for $\phi$ turns from sine Gordon into double-sine Gordon and in turn induces a mixing of the pseudo-modulus $\phi$ with the DW position modulus that results in wiggling of the DW that hosts the DW-skyrmion.
The ground state, however, corresponds to a DW that is rotated by $\tilde\vartheta$ in the $(x,y)$-plane for which the double-sine Gordon and the wiggling of the DW do not appear.
Since we are interested in the ground state here, we can without loss of generality set $\tilde\vartheta=0$ that corresponds simply to working with the Bloch DMI and the sine-Gordon Ansatz for $\phi$ in Eq.~\eqref{eq:phi_ansatz}.

\subsection{Numerical method}\label{sec:numerical_method}

We will utilize the so-called arrested Newton flow (aNF) algorithm, that
unlike gradient flow will accelerate towards the nearest local
``vacuum'' in field space (whereas gradient flow will decelerate
exponentially upon approach of the local ``vacuum'').
Of course, without the arrest, the Newton flow will blast right past
the local vacuum, so we monitor the static energy at every point in
the flow and arrest the fields by setting $\p_\tau\bn=0$ if the static
energy increases.
The flow equation can be written simply as
\beq
\p_\tau^2\bn = {\rm eom}_{\bn},
\eeq
with ${\rm eom}_{\bn}$ given in Eq.~\eqref{eq:eom} and spatial
derivatives are discretized with a five-point fourth-order stencil.
Due to the less important precision on the fictitious-time ($\tau$)
flow, we use a fast second-order discretization of the double
$\tau$-derivative.
The quadratic square grids that we use for the numerical computations
are typically of lattice sizes $1280^2$ with stepsizes smaller than
$0.04$. 

The aNF algorithm is much faster than gradient flow for obtaining
stable or metastable static solitons, which is in turn somewhat faster
than the Landau-Lifshitz-Gilbert (LLG) equation. In fact, the LLG
equation with large Gilbert damping coefficient is close to gradient
flow with a small symplectic flow component, whereas LLG flow with
small Gilbert damping is close to a loopy symplectic flow, which may
not efficiently arrive at even close-in-field-space local ``vacua'' (fixed points).

\section{Results}\label{sec:results}

With the initial configurations of Sec.~\ref{sec:initial_conf}, the
numerical method of Sec.~\ref{sec:numerical_method} in hand, we are
now ready to perform a vast array of numerical computations that we
present in this section.

\begin{table}[!htp]
  \begin{center}
    \phantom{.}\\[10pt]
    \begin{tabular}{r||c||c}
      &$\frac12B<K<\frac12B^2$ & $K>\frac12B^2$\\[3pt]
      \hline\hline
      DW-type & $\theta_L^-$ & $\theta_S^+$\\
      hosted stable soliton & bimeron ($+$) & bimeron ($-$)\\
      hosted excited soliton & anti-bimeron$^*$ ($-$) & anti-bimeron$^*$ ($+$)\\[3pt]
      \hline\hline
      excited DW-type & $\theta_S^+$ & $\theta_L^-$\\[3pt]
      \hline\hline
      anti-DW-type & $\theta_S^-$ & $\theta_S^-$\\
      hosted excited soliton & anti-skyrmion$^*$ ($+$) & anti-skyrmion ($+$)\\[3pt]
      \hline\hline
    \end{tabular}
    \caption{DW and anti-DW types depending on $K$ being larger or
      smaller than the degenerate point \eqref{eq:degenerate_point}
      (i.e.~$\frac12B^2$) as well as the topological solitons that
      they may host.
      The DWs host stable bimerons with a kink ($+$) and anti-kink
      ($-$) ansatz, whereas an excited state is the DW hosting
      anti-bimerons.
      The kink or anti-kink Ansatz depends again on $K$ being larger
      or smaller that the degenerate point.
      The opposite type of DW also exists, but are excited states.
      Finally, the anti-DW is throughout of type $\theta_S^-$ and as
      an excited state can host an anti-skyrmion with the kink Ansatz
      ($+$).
      Notice that the stable solitons are stable in part of the parameter
      space and those with ${}^{*}$ exist only in part of the
      parameter space.
    }
    \label{tab:DW_soliton_type}
  \end{center}
\end{table}

\begin{figure*}[!htp]
  \centering
  \mbox{\subfloat[DW hosted solitons]{\includegraphics[width=0.49\linewidth]{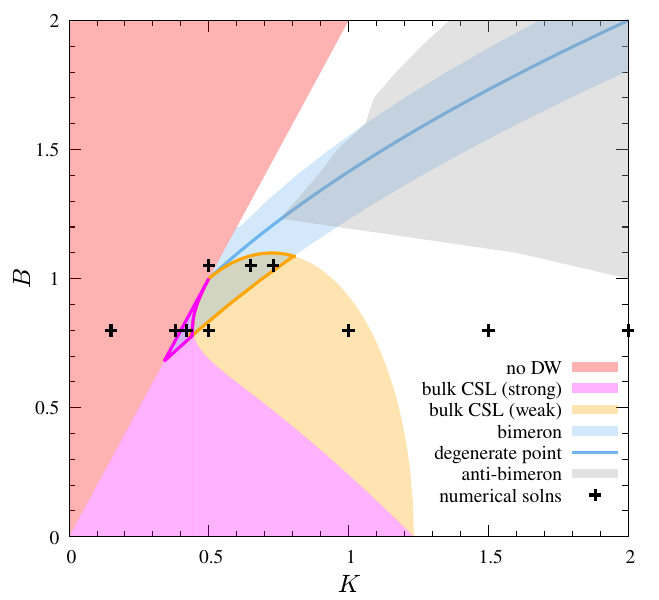}}
  \subfloat[skyrmion crystal $+$ bimeron crystal $+$ anti-DW (overlay)]{\includegraphics[width=0.49\linewidth]{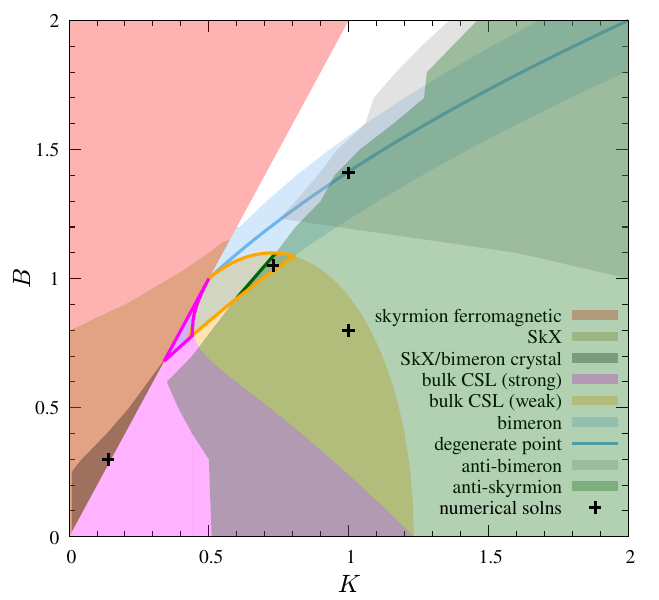}}}
  \caption{Phase diagram in $K$-$B$ parameter space.
    (a) The red region ($K<\frac12B$) contains no DWs.
    The blue region contains stable bimerons on DWs, whereas outside
    the blue region the bimerons are excited states and the empty DW
    is the stable state. 
    The blue solid line corresponds to the degenerate point
    ($K=\frac12B^2$).
    In the magenta region the ground state is the CSL, meaning that
    the DW-anti-DWs have a lower energy than the homogeneous vacua,
    whereas in the orange region the DW has a positive energy, but the
    sum of the anti-DW and DW's energies is negative, so the state is
    stable when averaged over space.
    The DW also contains an anti-bimeron, which is always an excited
    state and exists only in the gray-colored region.
    The overlap between the blue bimeron region and the magenta bulk CSL with the strong condition is marked with a magenta curve, whereas the overlap between the bimeron and the bulk CSL with the weak condition is marked with an orange curve. 
    (b) Shows the anti-DW's soliton as well as the skyrmion phase and
    bimeron phase (for $K<\frac12B$). In the green overlaid region the
    anti-skyrmion exists on the anti-DW as an excited state.
    The red region corresponds to the ferromagnetic skyrmion phase
    (excited state), whereas the olive colored region is the stable
    skyrmion lattice (SkX).
    The dark region below the skyrmion lattice is a bimeron lattice than can coexist with the skyrmion lattice.
    The overlap between the bulk CSL with the weak condition and the green region with the anti-skyrmion is marked with a green curve.
  }
  \label{fig:phasediagram}
\end{figure*}

We start by presenting the $K$-$B$ phase diagram in
Fig.~\ref{fig:phasediagram} and details about the (anti-)DW solutions
and the solitons in Tab.~\ref{tab:DW_soliton_type}, viz.~which DW or
anti-DW solution and which kink or anti-kink Ansatz give rise to which
kind of soliton.

Fig.~\ref{fig:phasediagram}(a) shows the DW-part of the $K$-$B$
phase diagram and the solitons that are hosted on the DW.
First of all, the DW requires $K>\frac12B$ (see
Eqs.~\eqref{eq:vac1}-\eqref{eq:vac2}) for its existence, that is,
without a large-enough in-plane anisotropy there will not exist two
degenerate vacua and hence no DW.
First, as described in Tab.~\ref{tab:DW_soliton_type}, the DW-type
depends on $K$ being smaller or larger than the degenerate point
\eqref{eq:degenerate_point} (i.e.~$\frac12B^2$ marked with a solid
blue line in Fig.~\ref{fig:phasediagram}).
In particular, for $K>\frac12B^2$ and for $B<1$, $K>\frac12B$, the
DW-type is $\theta_S^+$, whereas for $\frac12B<K<\frac12B^2$ the
DW-type is $\theta_L^-$.
Both DW types can host a bimeron that can be understood as a confined
half-skyrmion, but on the $\theta_S^+$ DW the anti-kink (i.e.~a minus
sign in Eq.~\eqref{eq:phi_ansatz}) gives rise to the bimeron, whereas
on the $\theta_L^-$ DW the kink (i.e.~a plus sign in
Eq.~\ref{eq:phi_ansatz}) gives rise to the bimeron (see
Tab.~\ref{tab:DW_soliton_type}).
The bimeron exists everywhere in the displayed parameter space (for
$K>\frac12B$), but is only the ground state in the blue region,
meaning that its energy is lower than the empty DW's energy.
The DW and anti-DW lattice is, however, not everywhere the ground
state, but with the strong condition \eqref{eq:CSL_strong_cond} this
lattice \emph{is} the ground state and this region is displayed in
Fig.~\ref{fig:phasediagram}(a) with a magenta colored region.
The weak condition \eqref{eq:CSL_weak_cond} for the CSL phase being
the ground state is shown in the figure with an orange color.
The weak condition implies that whereas the DW's energy is positive,
the anti-DW's energy is so negative that the sum of the two is
negative and hence the phase has a lower energy than the homogeneous
phase, when averaged over space.

\begin{figure*}[!htp]
  \centering
  \mbox{\subfloat[$B=0$]{\includegraphics[width=0.33\linewidth]{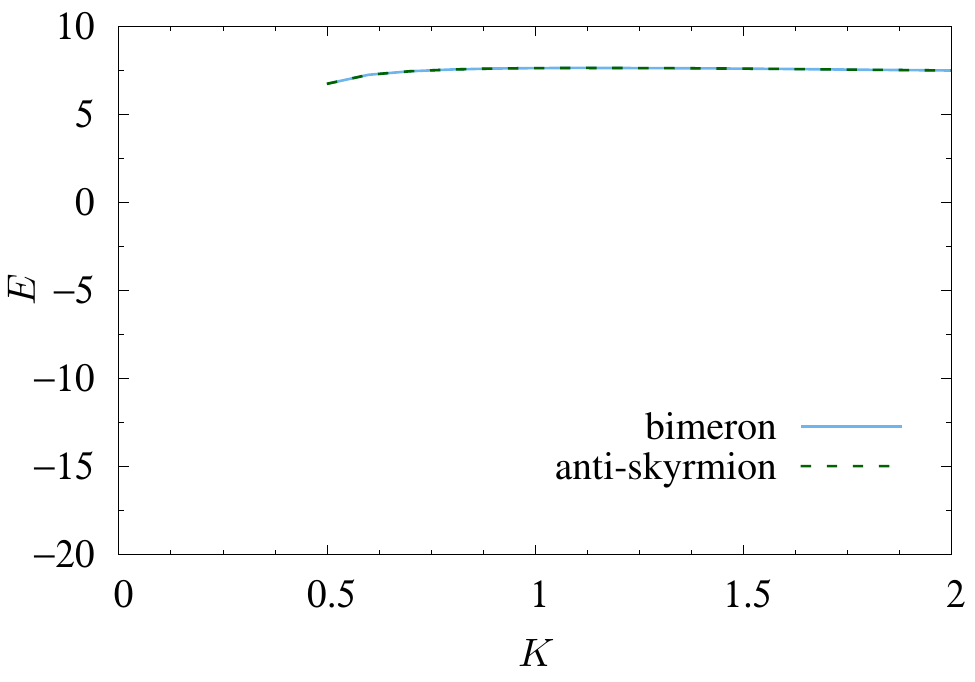}}
    \subfloat[$B=0.5$]{\includegraphics[width=0.33\linewidth]{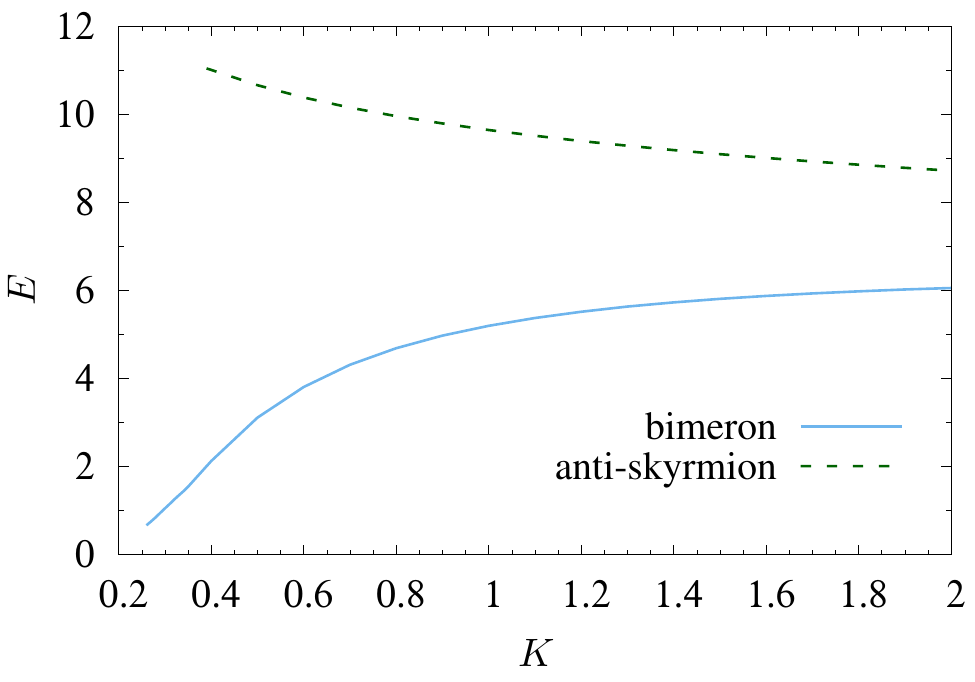}}
    \subfloat[$B=0.7$]{\includegraphics[width=0.33\linewidth]{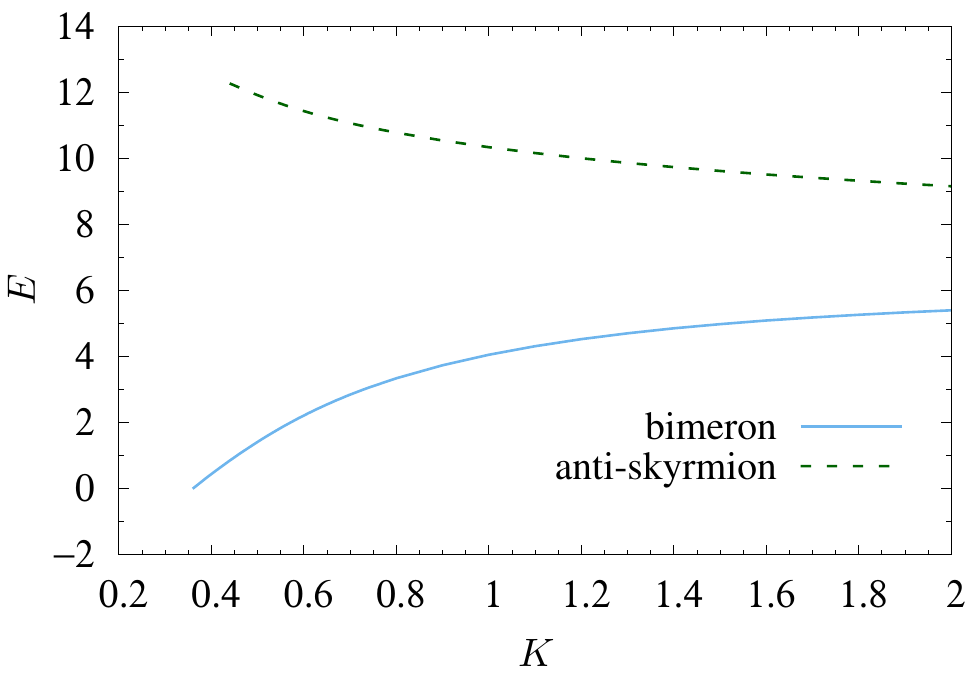}}}
  \mbox{\subfloat[$B=0.8$]{\includegraphics[width=0.33\linewidth]{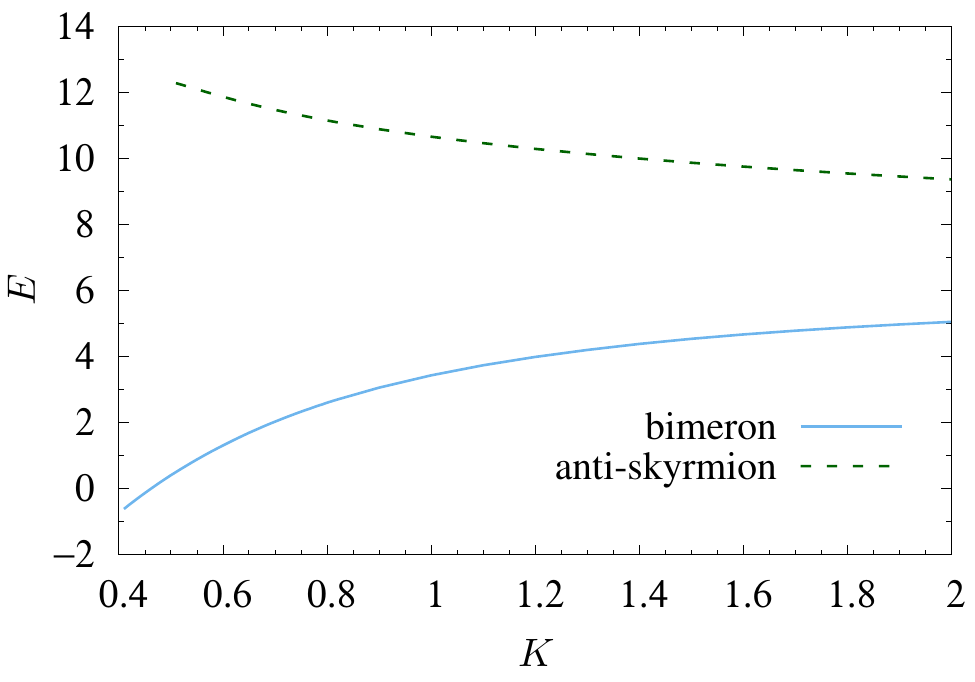}}
    \subfloat[$B=0.9$]{\includegraphics[width=0.33\linewidth]{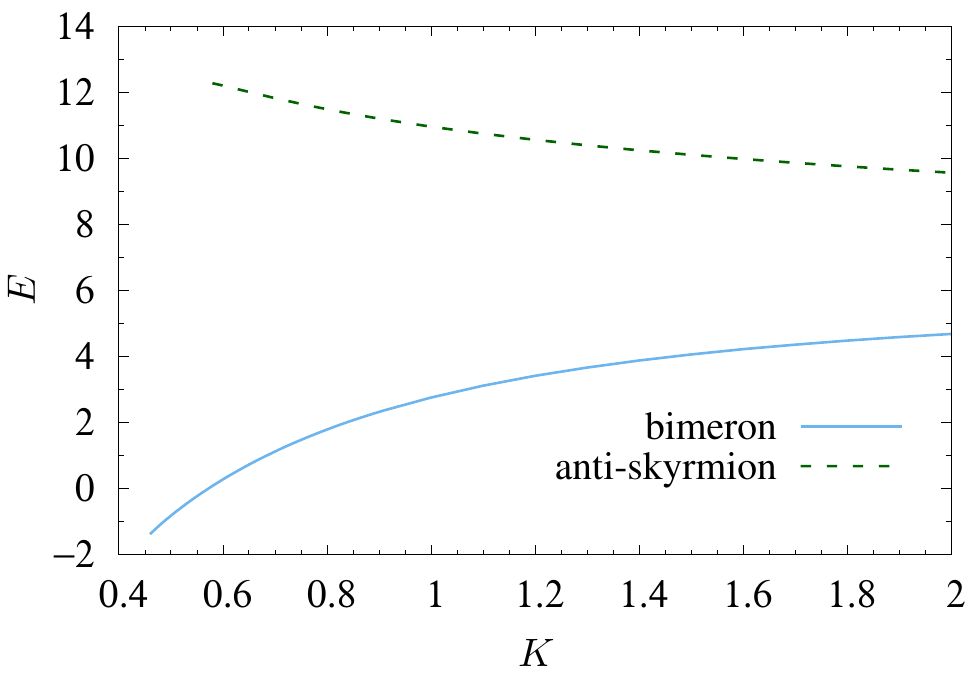}}
    \subfloat[$B=1$]{\includegraphics[width=0.33\linewidth]{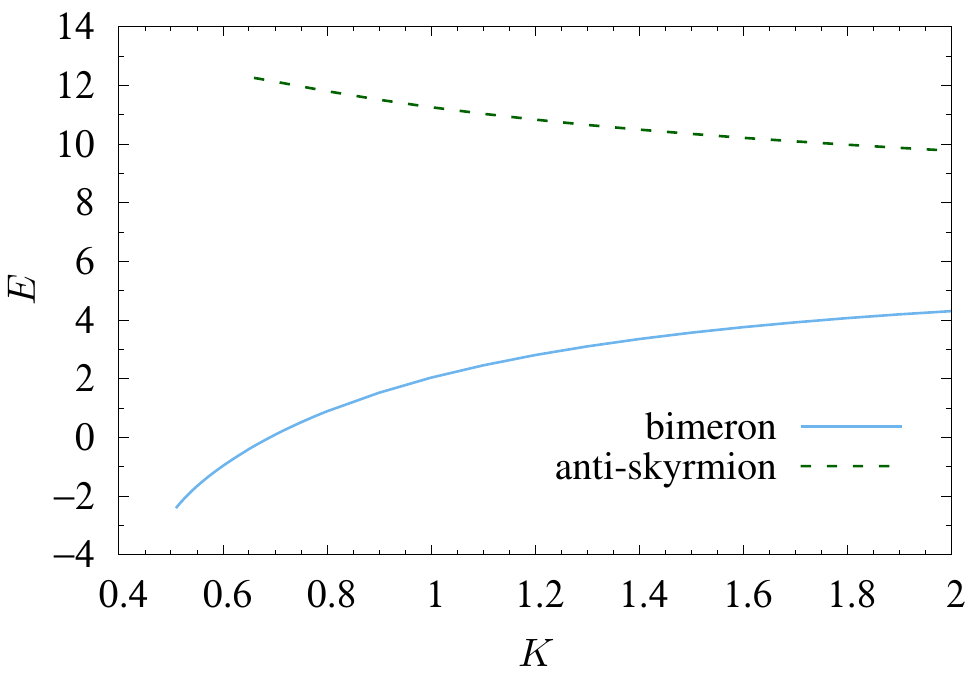}}}
  \mbox{\subfloat[$B=1.5$]{\includegraphics[width=0.33\linewidth]{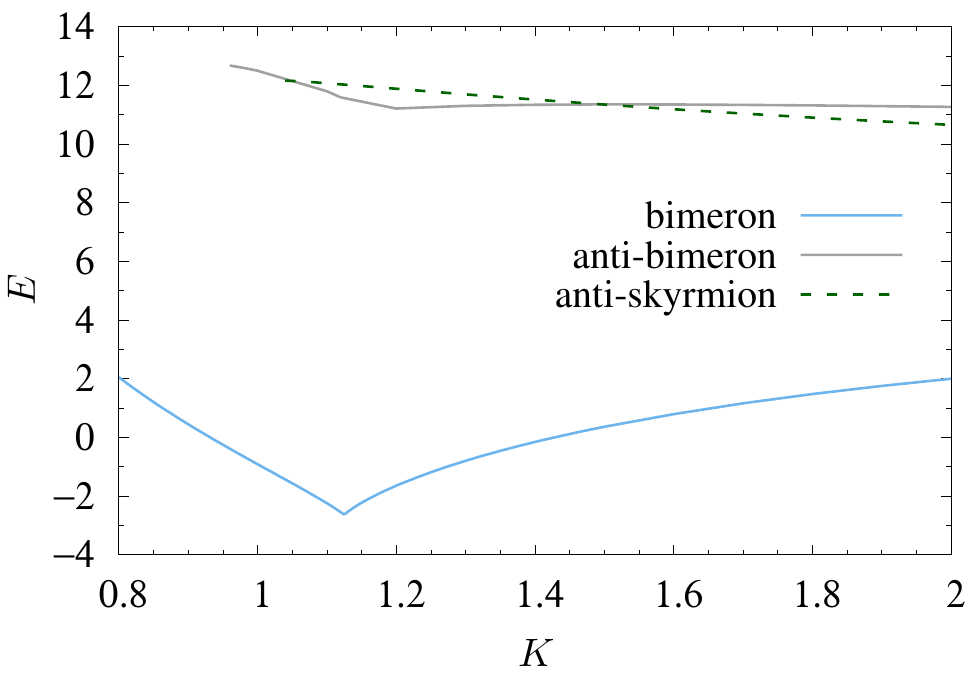}}
    \subfloat[$B=1.9$]{\includegraphics[width=0.33\linewidth]{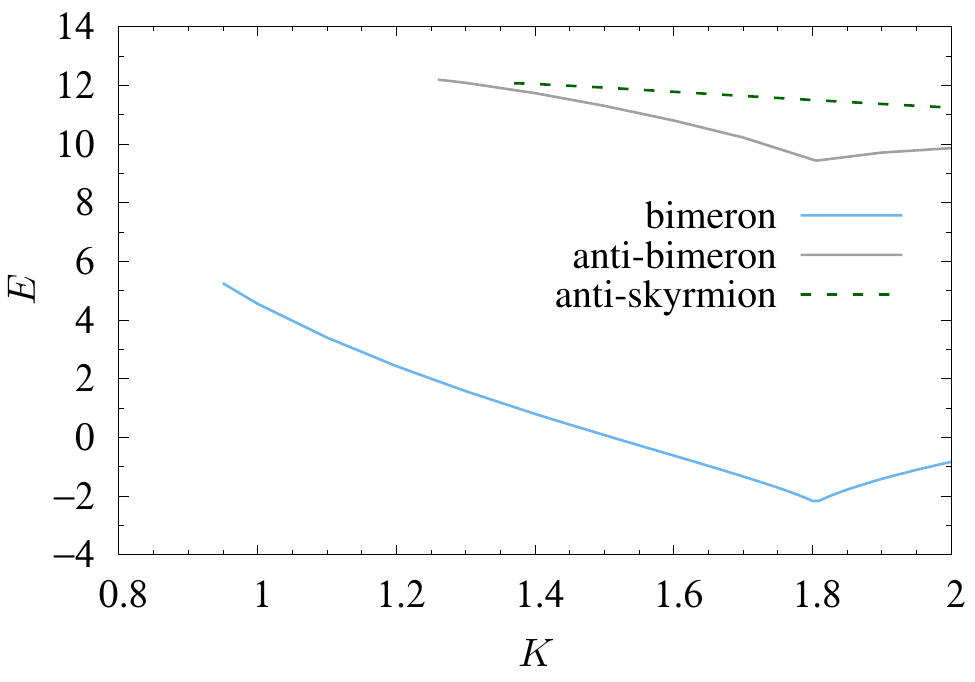}}
    \subfloat[$B=2$]{\includegraphics[width=0.33\linewidth]{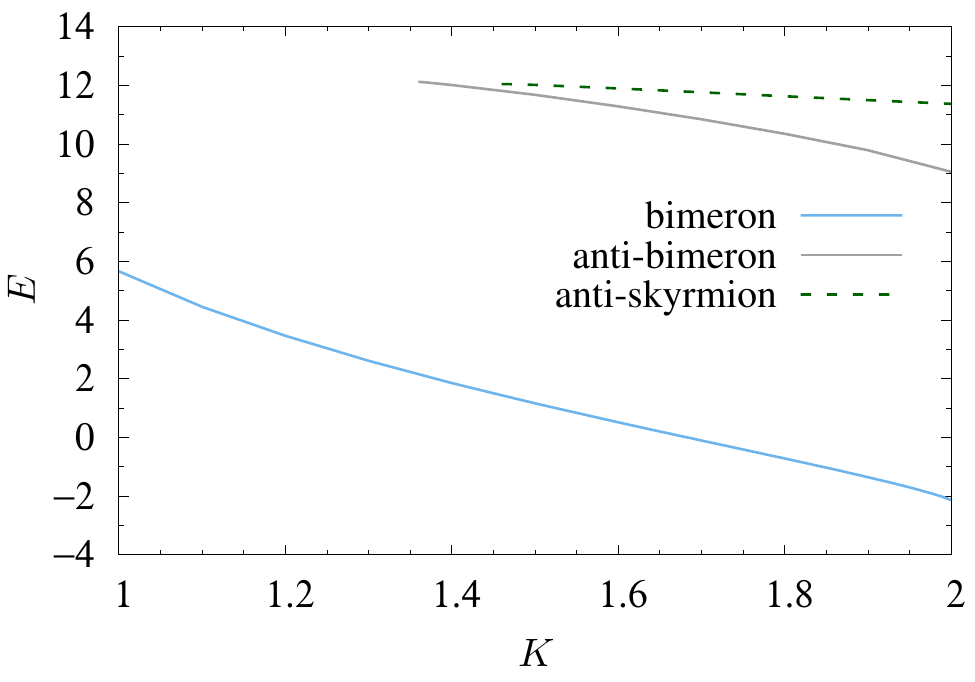}}
  \hspace{0.33\linewidth}}
  \caption{Bimeron and anti-bimeron energies at fixed slices of $B$ in
  the phase diagram (Fig.~\ref{fig:phasediagram}) as functions of $K$,
  measured relative to the DW's energy.
  The figures also show the anti-skyrmion energies, but relative to
  the anti-DW's energy as they live on the anti-DW.
  In panel (a), the bimeron and anti-skyrmion's energies are
  degenerate.
  Negative bimeron energies make it the ground state, whereas the
  positive energy states are excited states of the DW. 
  The colors of the curves match the soliton type in Fig.~\ref{fig:phasediagram}.
  }
  \label{fig:energies}
\end{figure*}

The CSL on CSL phase corresponds to the overlap between the CSL bulk
phase and the stable bimeron phase, which is the overlap between the
magenta region and the blue region in Fig.~\ref{fig:phasediagram}(a)
in case of the strong CSL condition
(i.e.~Eq.~\eqref{eq:CSL_strong_cond}) and the overlap between the
union of the orange and the magenta region with blue region in
Fig.~\ref{fig:phasediagram}(a) in case of the weak CSL condition
(i.e.~Eq.~\eqref{eq:CSL_weak_cond}). 
The energies of the bimeron with respect to the DW's energy are shown
for slices of fixed $B$ as functions of $K$ in
Fig.~\ref{fig:energies}; this figure also shows the energies of the
anti-bimerons that are excited states on the DW (again relative to
the DW's energy) and also the energies of the anti-skyrmions, however,
relative to the anti-DW's energy (as they live on the anti-DW).
When the bimeron's energy is negative in Fig.~\ref{fig:energies} then
the DW hosting the bimeron is the ground state (called DW-skyrmion in
Ref.~\cite{Amari:2024jxx}).
One can see from Fig.~\ref{fig:energies}(c) that the on-set of the
stable bimeron phase begins around $B=0.7$ for $K\approx\frac12B$
(viz.~smallest possible $K$).
In Fig.~\ref{fig:energies} it is not possible to see the bulk CSL
conditions \eqref{eq:CSL_strong_cond} and \eqref{eq:CSL_weak_cond},
but from Fig.~\ref{fig:phasediagram}(a) they require the smallest
possible $K$ which is why the slices $B=0.7,0.8,0.9,1$ are shown in
Fig.~\ref{fig:energies}.


\begin{figure*}[!htp]
  \centering
  \includegraphics[width=0.216\linewidth]{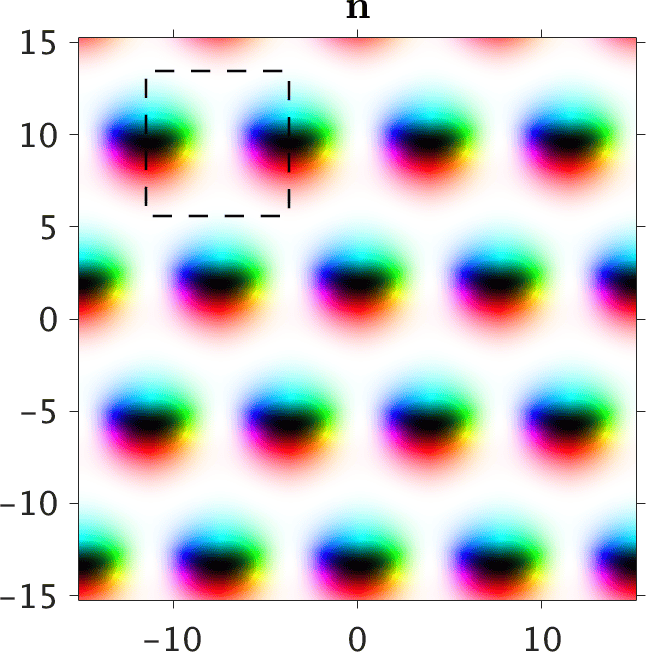}\hfill
  \includegraphics[width=0.25\linewidth]{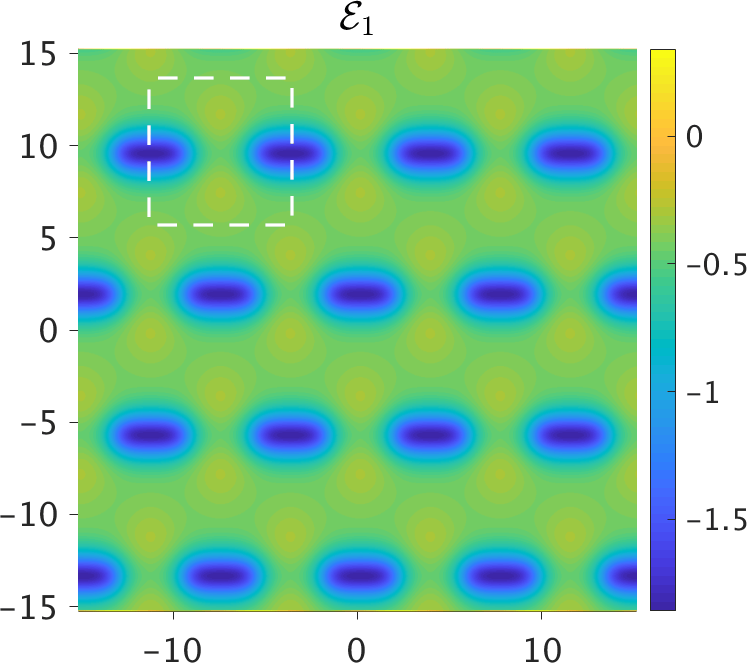}\hfill
  \includegraphics[width=0.25\linewidth]{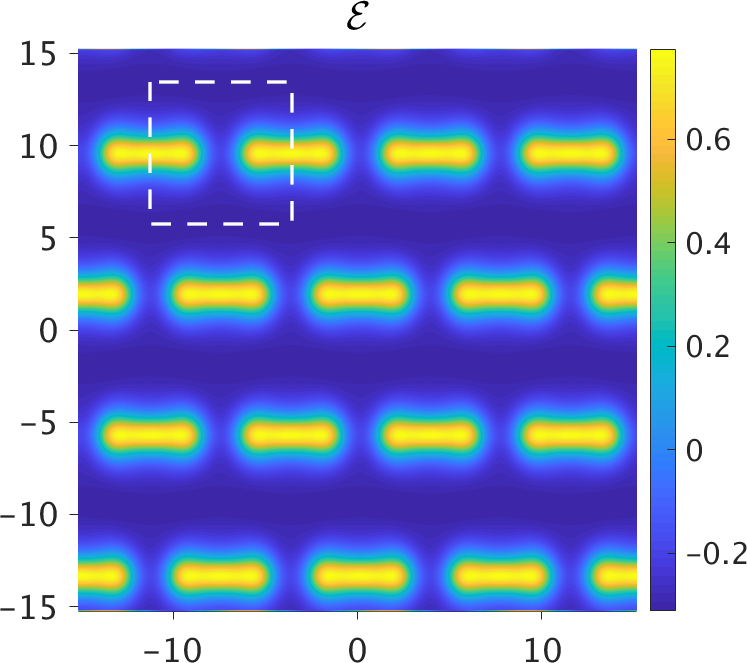}\hfill
  \includegraphics[width=0.251\linewidth]{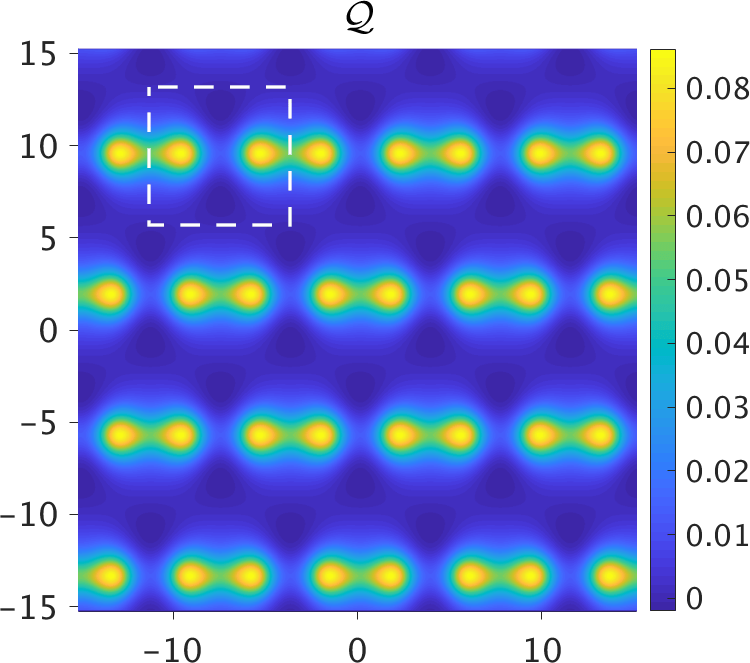}
  \caption{CSL on CSL phase (strong condition): $K=0.42$, $B=0.8$.
    The four panels display the magnetization vector $\bn$, the DMI
    energy density $\calE_1$, the total energy density $\calE$ and
    finally the topological charge density $\calQ$.
    The coloring scheme of the magnetization vector: 
    $\bn=\be_3$ is white, $\bn=-\be_3$ is black, $\bn=\be_1$ is red, 
    $\bn=\be_2$ is light green, $\bn=-\be_1$ is cyan and $\bn=-\be_2$ is violet. 
    The energy of the solution in this figure is minimized with
    respect to the dimensions in both the $x$- and the $y$-direction
    subject to periodic boundary conditions.
    A single bimeron is marked with a dashed line in all panels.
  }
  \label{fig:K0.42B0.8}
\end{figure*}


\begin{figure*}[!htp]
  \centering
  \includegraphics[width=0.218\linewidth]{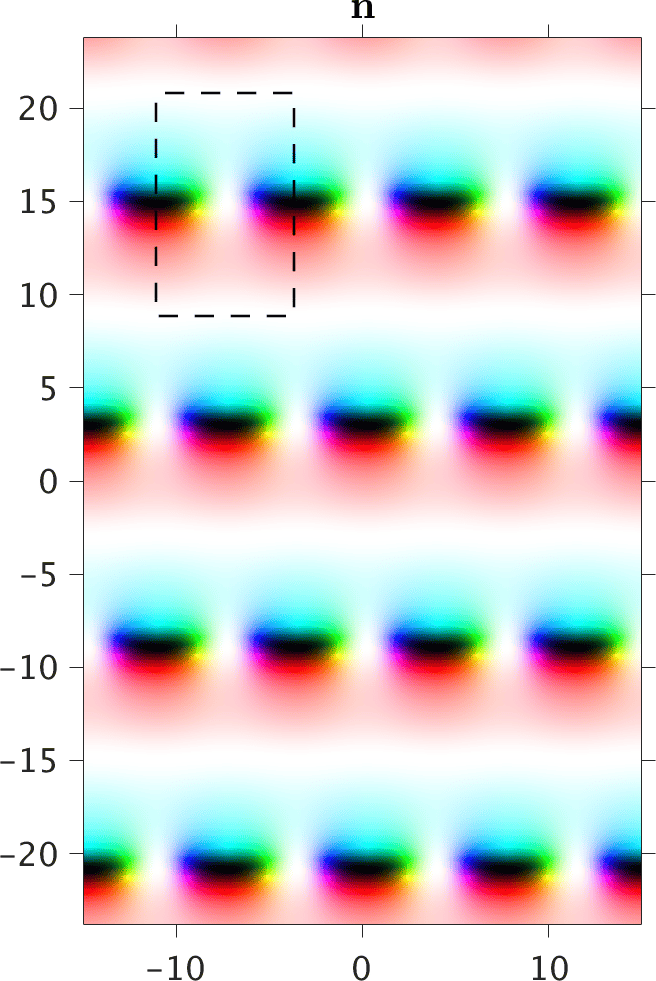}\hfill
  \includegraphics[width=0.251\linewidth]{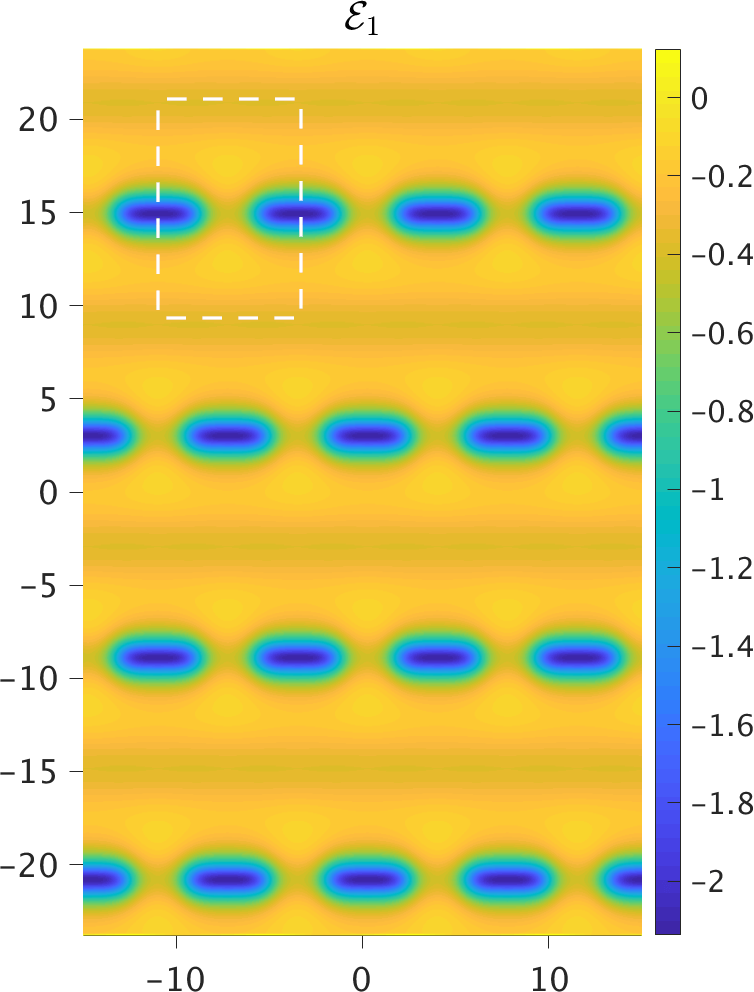}\hfill
  \includegraphics[width=0.25\linewidth]{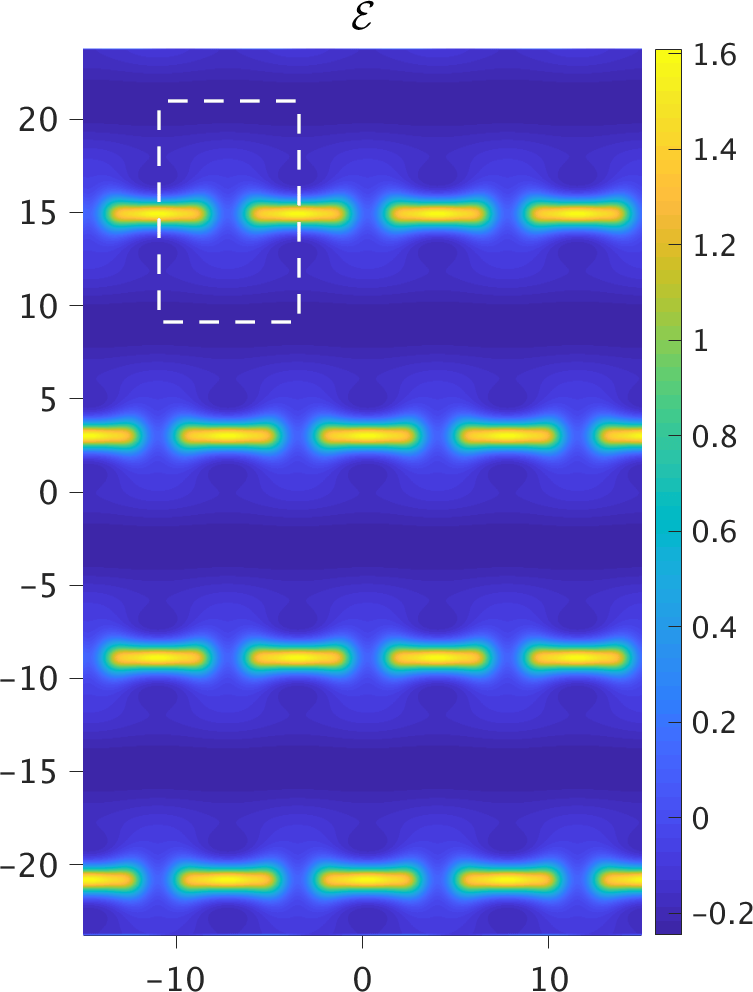}\hfill
  \includegraphics[width=0.251\linewidth]{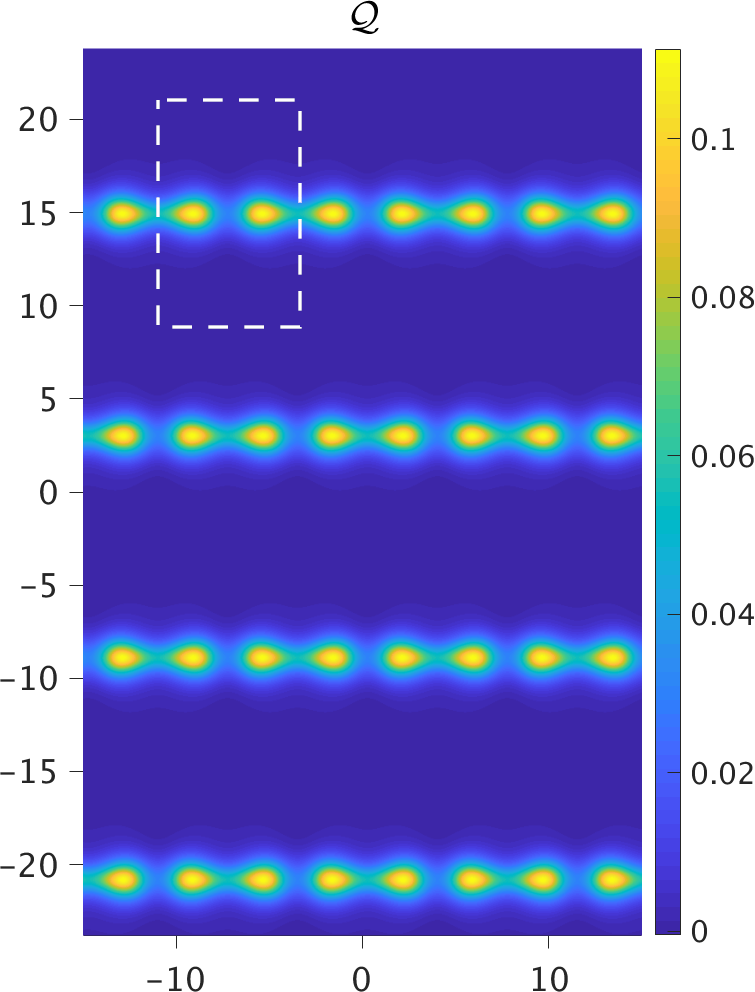}
  \put(-510,160){(a)}\\
  \includegraphics[width=0.217\linewidth]{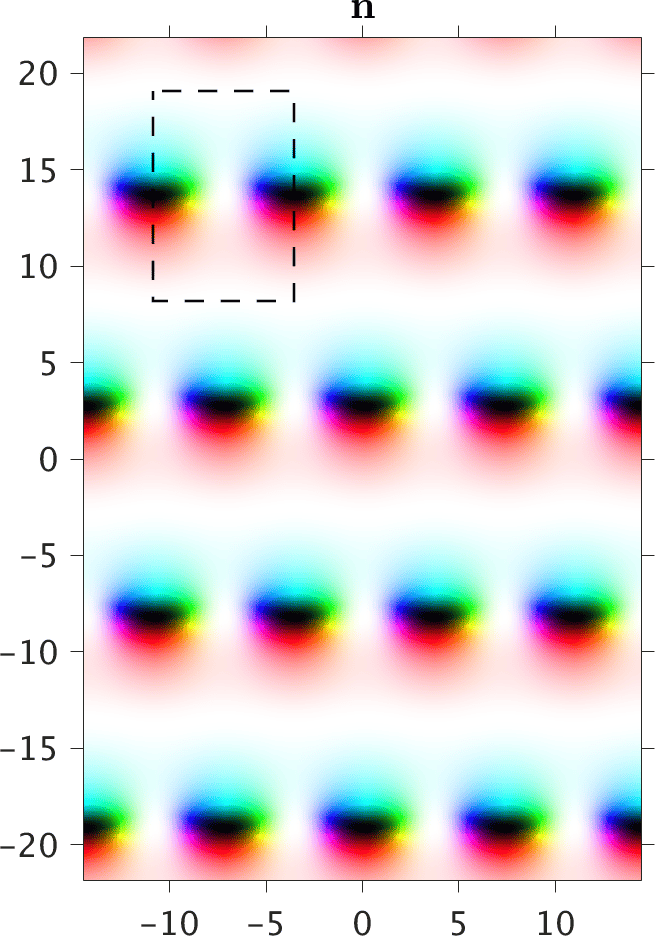}\hfill
  \includegraphics[width=0.25\linewidth]{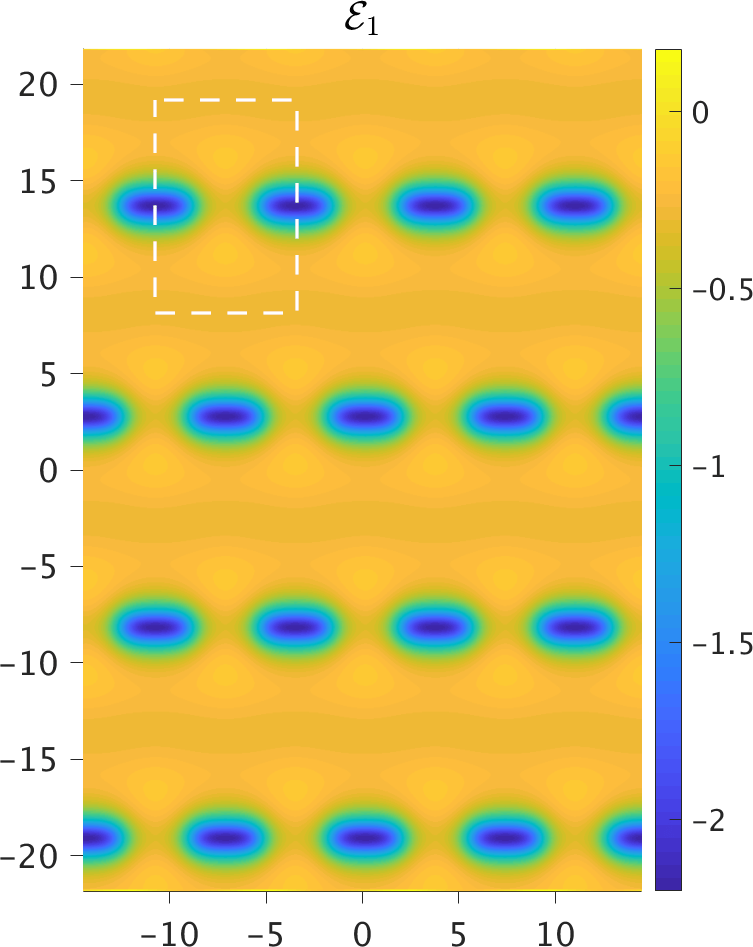}\hfill
  \includegraphics[width=0.25\linewidth]{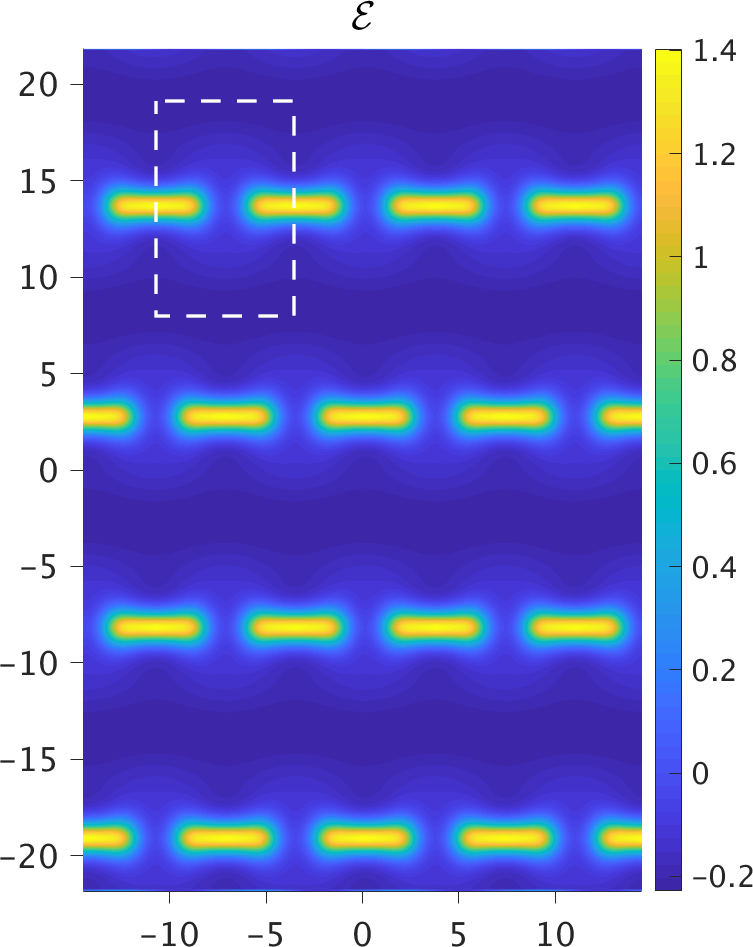}\hfill
  \includegraphics[width=0.251\linewidth]{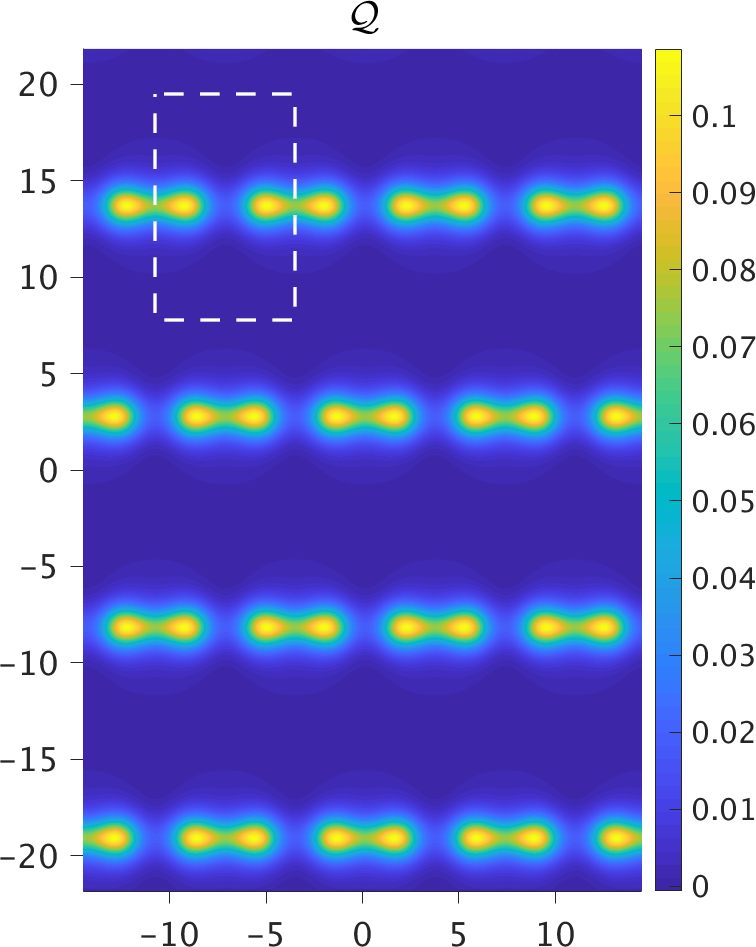}
  \put(-510,156){(b)}
  \caption{CSL on CSL phase (weak condition): (a) $K=0.73$ and (b)
    $K=0.65$, $B=1.05$ for both.
    The energies of the solutions in this figure are minimized with
    respect to the dimensions in both the $x$- and the $y$-direction
    subject to periodic boundary conditions.
    A single bimeron is marked with a dashed line in all panels.
    For details of the four columns of panels, see the caption of
    Fig.~\ref{fig:K0.42B0.8}.
  }
  \label{fig:K0.73_0.65B1.05}
\end{figure*}

The CSL on CSL solution with the strong condition
(Eq.~\eqref{eq:CSL_strong_cond}) is shown in Fig.~\ref{fig:K0.42B0.8}
and with the weak condition (Eq.~\eqref{eq:CSL_weak_cond}) is shown in
Fig.~\ref{fig:K0.73_0.65B1.05} and for clarity the bimeron is marked
with a dashed square in all panels.
The strong condition's solution in Fig.~\ref{fig:K0.42B0.8} is the
ground state for $K=0.42$ and $B=0.8$ and is locally stable, meaning
that the energy density of both the DW and the anti-DW are negative
compared with the homogeneous vacuum state.
The weak condition's solutions shown in
Fig.~\ref{fig:K0.73_0.65B1.05}(a)[(b)] for $K=0.73$ [$K=0.65$] and
$B=1.05$ is only globally a ground state, i.e.~when averaged over
space: the DW's energy density is positive and the anti-DW's energy is
negative, but the sum is negative too making this state energetically
favorable over the homogeneous vacuum state.
Two observations can be made from the two figures, the strong
condition's solution in Fig.~\ref{fig:K0.42B0.8} has a smaller
DW-anti-DW separation (in the $y$-direction) than the weak condition's
solutions in Fig.~\ref{fig:K0.73_0.65B1.05}: this is due to the larger
value of $B$ in the latter.
The other observation is that the DW's length (in the $x$-direction)
is smaller for the strong condition's solution in
Fig.~\ref{fig:K0.42B0.8} than for the weak condition's solution in
Fig.~\ref{fig:K0.73_0.65B1.05}(a) (the length in the $x$-direction of the
black areas in the left-most panel labeled $\bn$): this is due to the
larger value of $K$ in the latter.
For even larger values of $K$ the DW's length increases further and it
becomes much clearer that it can indeed be interpreted as a DW and not
just a deformed skyrmion (see below).

\begin{figure*}[!htp]
  \centering
  \includegraphics[width=0.216\linewidth]{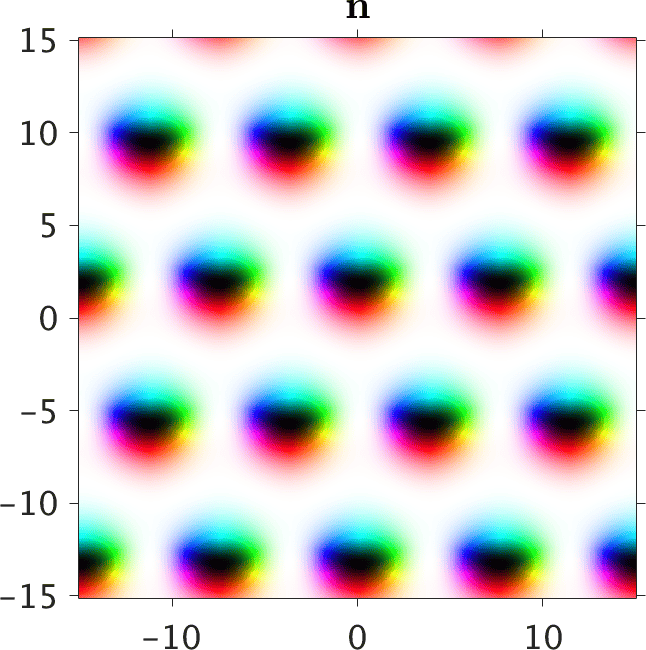}\hfill
  \includegraphics[width=0.25\linewidth]{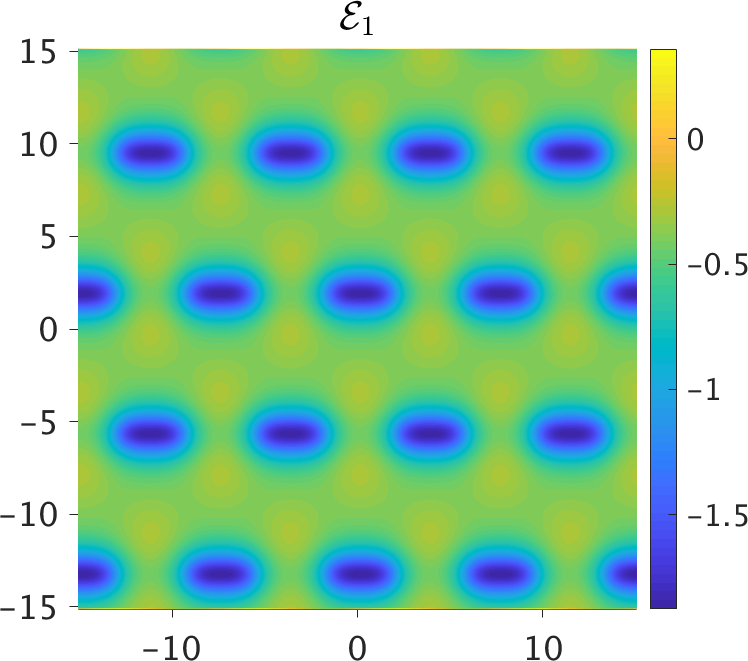}\hfill
  \includegraphics[width=0.25\linewidth]{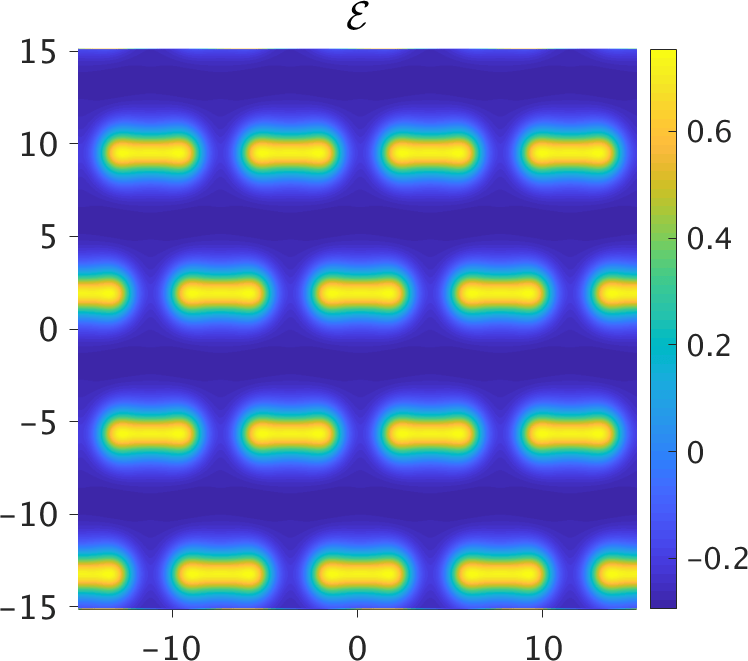}\hfill
  \includegraphics[width=0.251\linewidth]{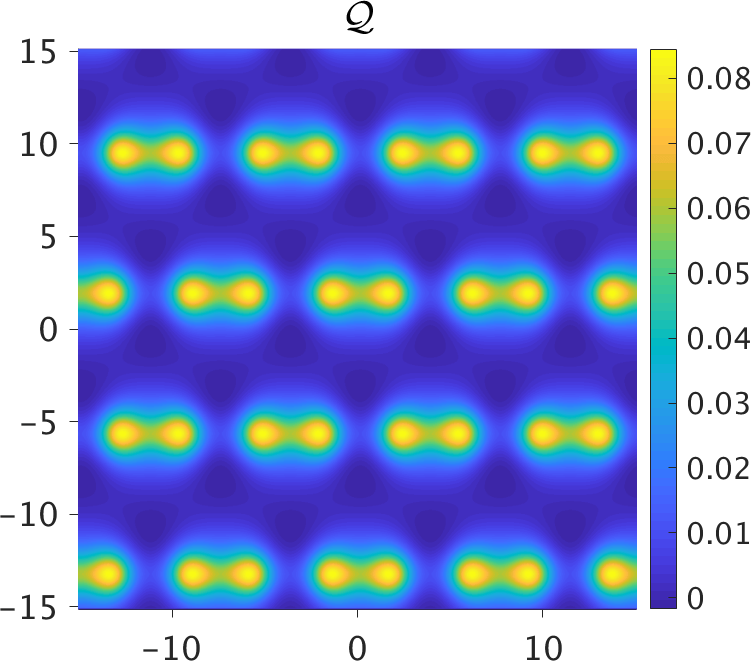}
  \put(-510,110){(a)}\\
  \includegraphics[width=0.217\linewidth]{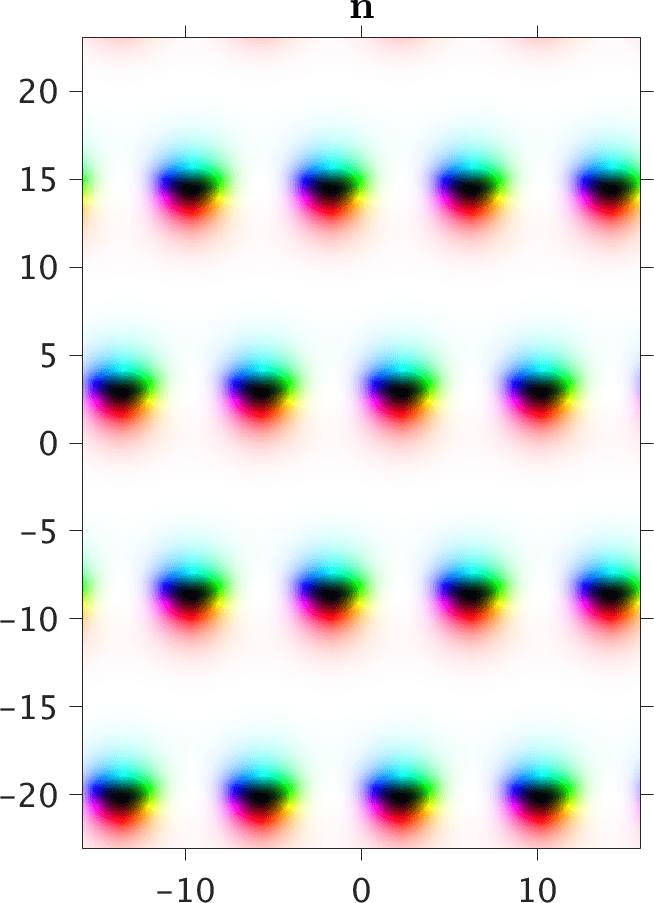}\hfill
  \includegraphics[width=0.25\linewidth]{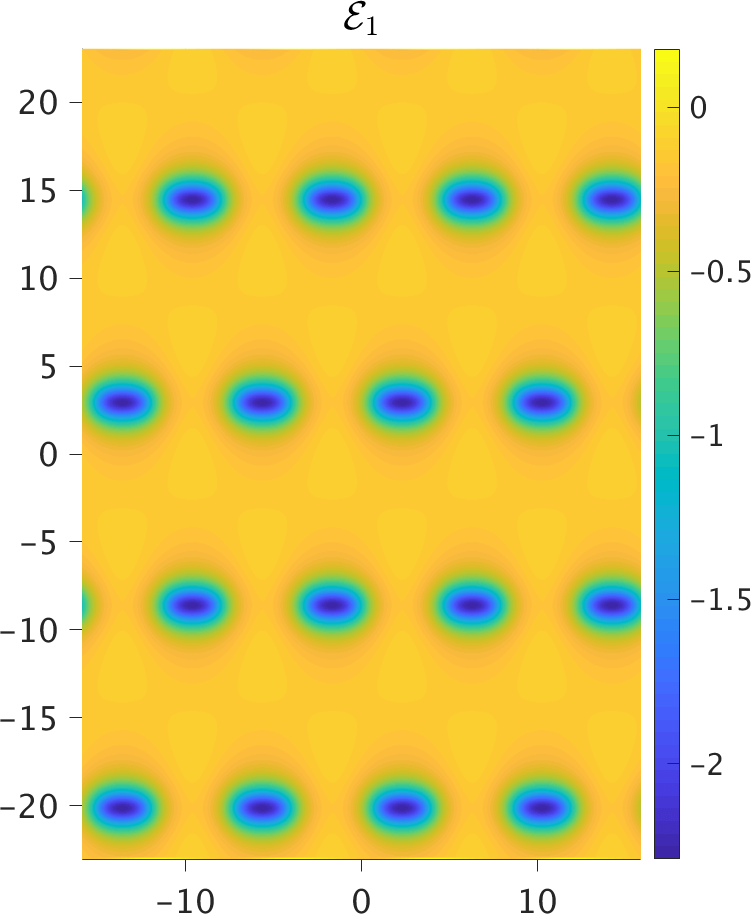}\hfill
  \includegraphics[width=0.244\linewidth]{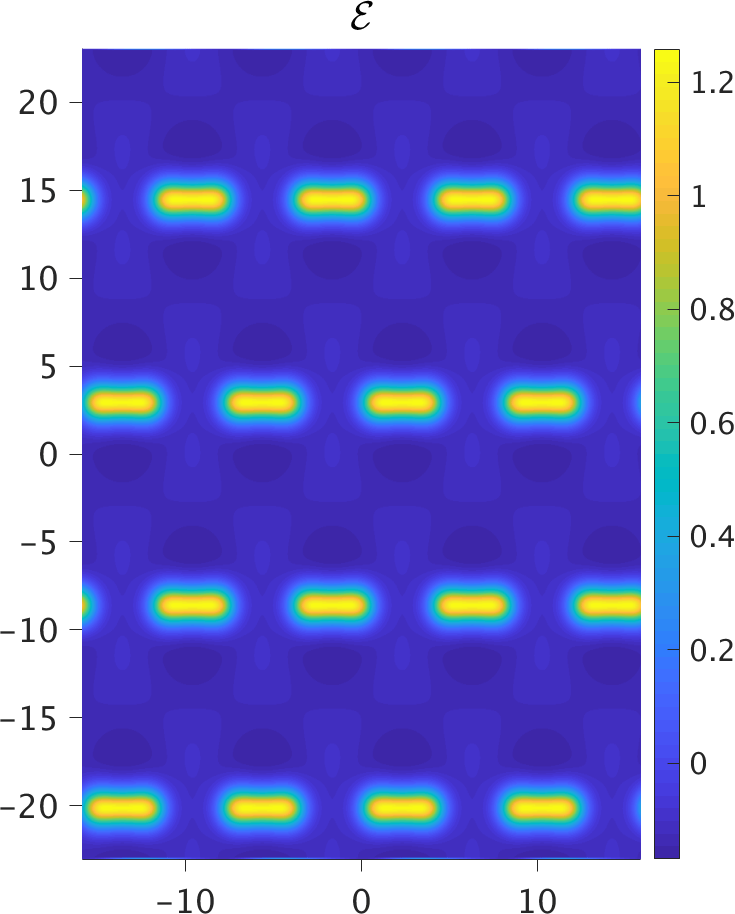}\hfill
  \includegraphics[width=0.251\linewidth]{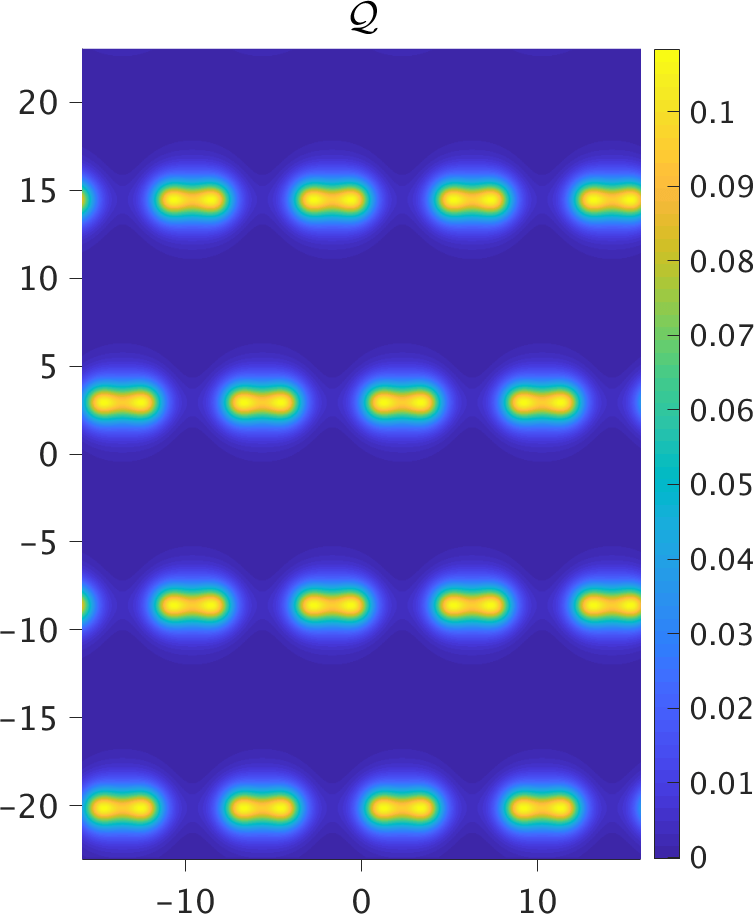}
  \put(-510,150){(b)}
  \caption{Skyrmion lattice phase (stretched hexahonal lattice): (a) $K=0.38$, $B=0.8$, and (b)
    $K=0.5$, $B=1.05$.
    The energies of the solutions in this figure are minimized with
    respect to the dimensions in both the $x$- and the $y$-direction
    subject to periodic boundary conditions.
    For details of the four panels, see the caption of
    Fig.~\ref{fig:K0.42B0.8}.
  }
  \label{fig:K0.38B0.8}
\end{figure*}

\begin{figure*}[!htp]
  \centering
  \includegraphics[width=0.216\linewidth]{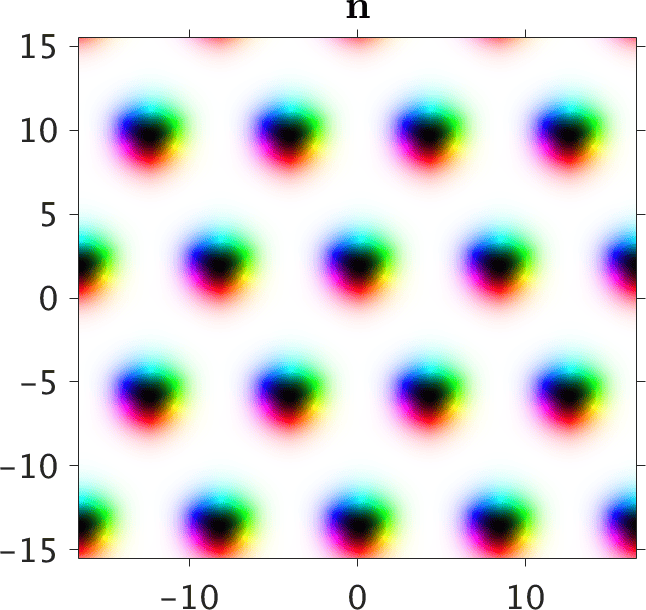}\hfill
  \includegraphics[width=0.25\linewidth]{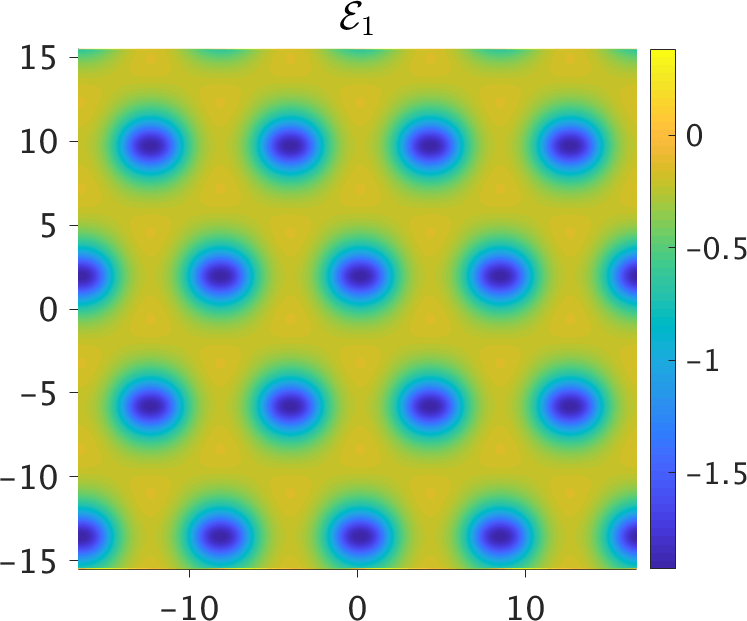}\hfill
  \includegraphics[width=0.245\linewidth]{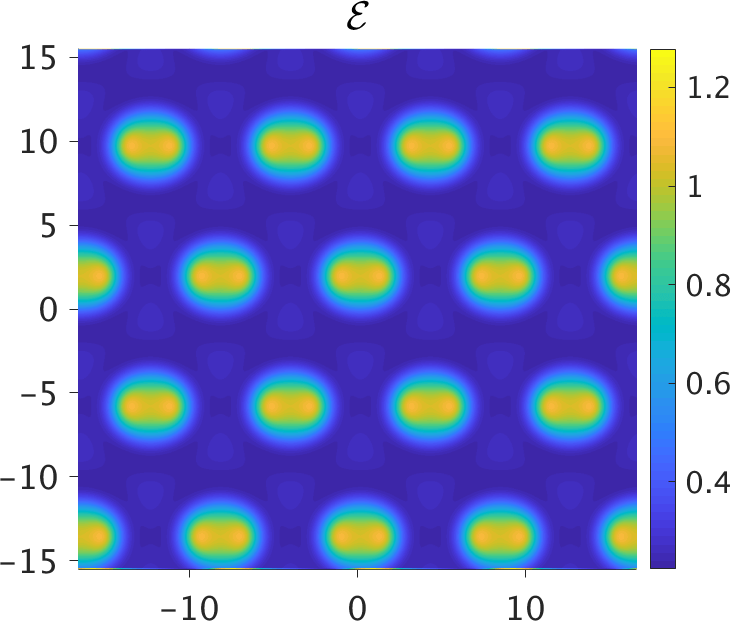}\hfill
  \includegraphics[width=0.251\linewidth]{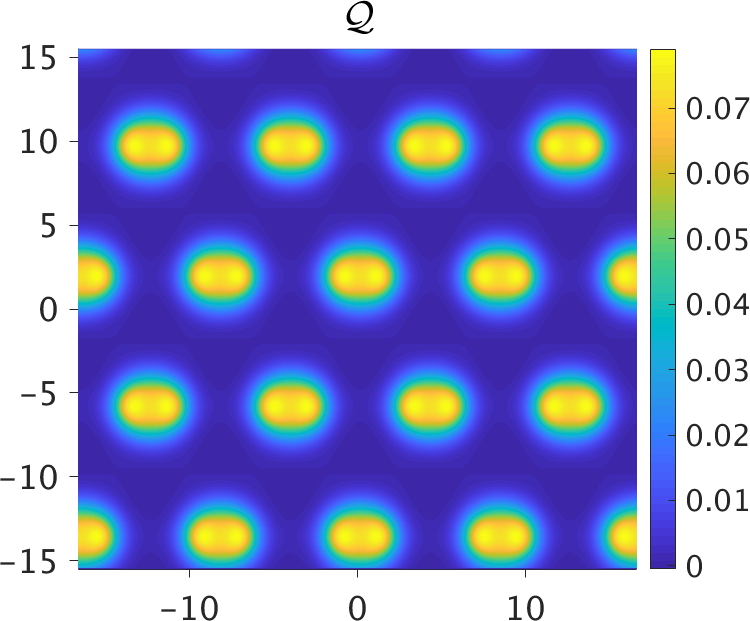}
  \caption{Skyrmion lattice phase: $K=0.15$, $B=0.8$.
    For details of the four panels, see the caption of
    Fig.~\ref{fig:K0.42B0.8}.
  }
  \label{fig:K0.15B0.8}
\end{figure*}

It is conceptually very interesting that the ground state on the DW is
a kink lattice and the DW-anti-DW periodic lattice is the ground state
in the direction orthogonal to the (anti-)DWs (i.e.~the
$y$-direction), which means that the ground state is a kink lattice
with a kink lattice hosted on the latter or for short: a CSL on a CSL.
This happens e.g.~for the strong condition
(Eq.~\eqref{eq:CSL_strong_cond}) for $K=0.42$ and $B=0.8$, see
Fig.~\ref{fig:K0.42B0.8}. 
However, the fact that $K$ is so close to $\frac12B$, which for
$B=0.8$ is just $0.4$ means that the CSL on a CSL is very close to the
skyrmion lattice phase, see Fig.~\ref{fig:phasediagram}.
Indeed, comparing the CSL on a CSL of Fig.~\ref{fig:K0.42B0.8} with
the skyrmion lattice of Fig.~\ref{fig:K0.38B0.8} for which $K=0.38$
(and $B=0.8$),
there is not visually a large difference: the short DWs (black
regions) are now interpreted as deformed skyrmions with the
deformation being due to the in-plane anisotropy term in the
potential.
Decreasing the anisotropy coefficient $K$ from $K=0.38$ (i.e.~smaller
than $\frac12B=0.4$) to $K=0.15$, the deformation of the skyrmions is
now so small that the solution is similar to a standard hexagonal
lattice of skyrmions, see Fig.~\ref{fig:K0.15B0.8}.
The small value, $K=0.15$, now makes the skyrmion lattice hexagonal.
Similarly, there is not visually a large difference between the CSL on
a CSL with the weak condition in Fig.~\ref{fig:K0.73_0.65B1.05}(b) for
which $K=0.65$ ($B=1.05$) and the skyrmion lattice phase shown in
Fig.~\ref{fig:K0.38B0.8}(b) for which $K=0.5$ ($B=1.05$).
We can thus say that the transition from the CSL on a CSL to the
skyrmion lattice phase is a smooth crossover. 


\begin{figure*}[!htp]
  \centering
  \includegraphics[width=0.216\linewidth]{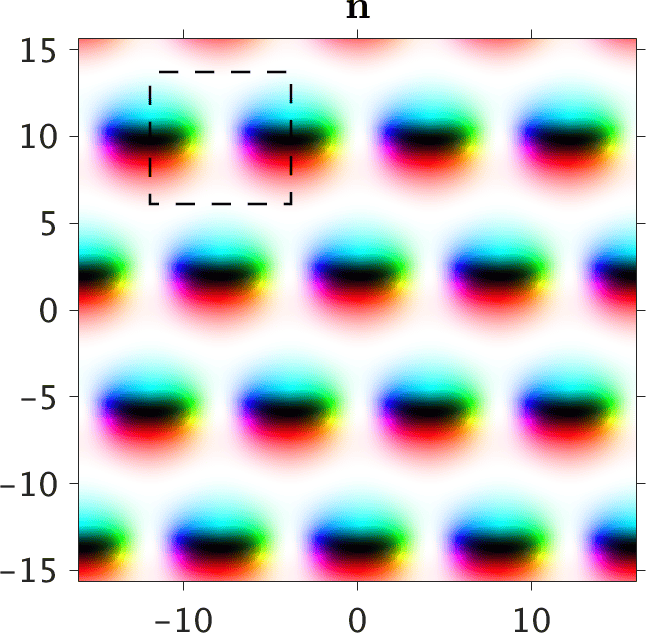}\hfill
  \includegraphics[width=0.25\linewidth]{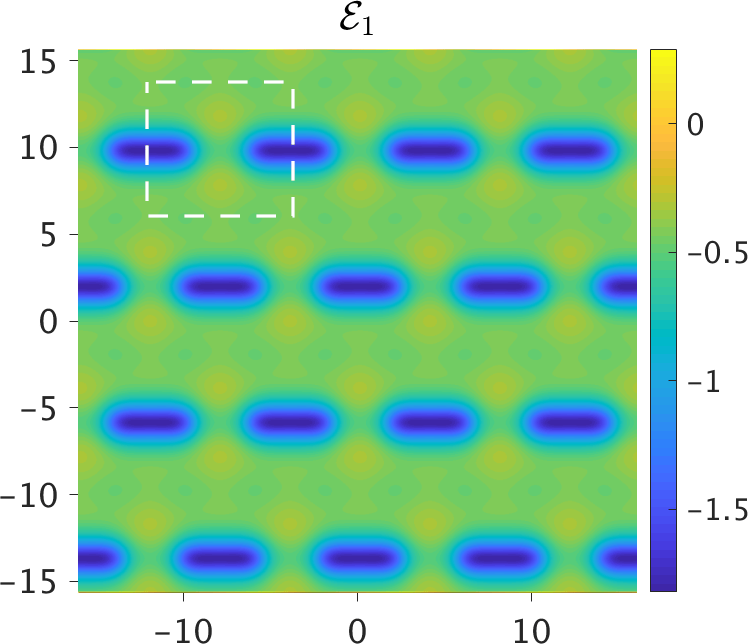}\hfill
  \includegraphics[width=0.25\linewidth]{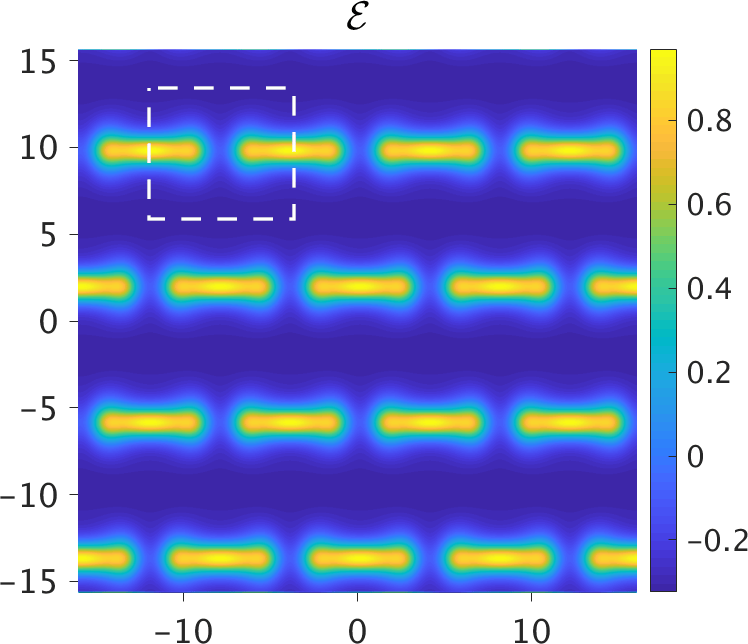}\hfill
  \includegraphics[width=0.251\linewidth]{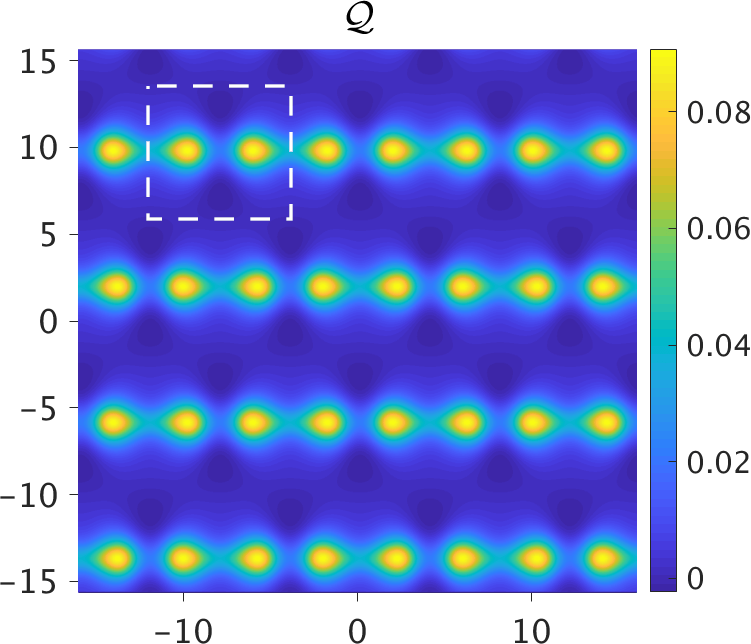}
  \put(-510,108){(a)}\\
  \includegraphics[width=0.216\linewidth]{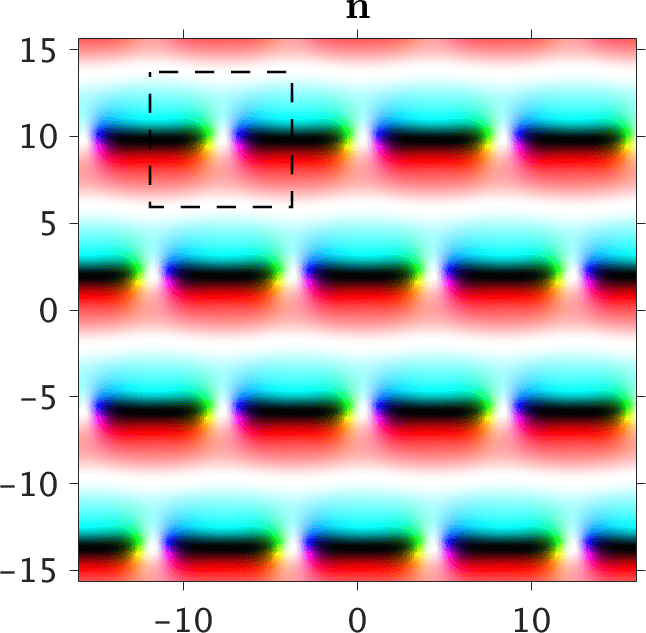}\hfill
  \includegraphics[width=0.25\linewidth]{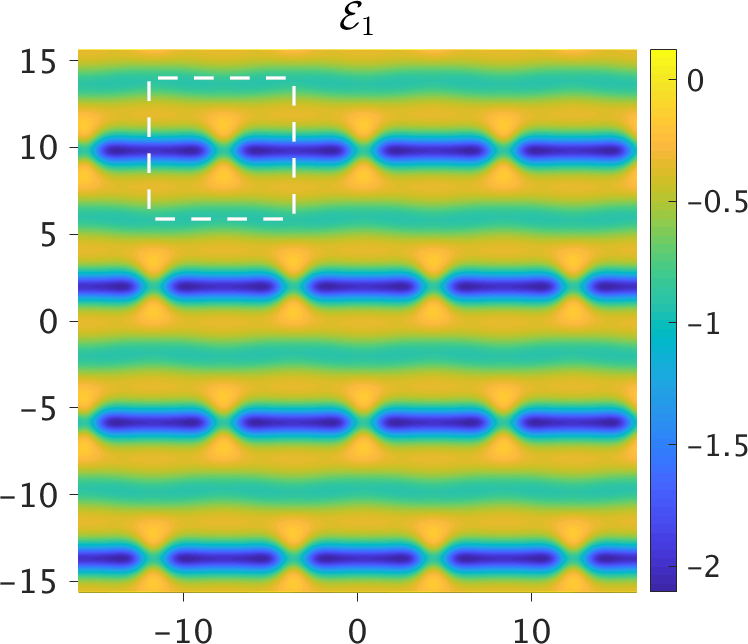}\hfill
  \includegraphics[width=0.244\linewidth]{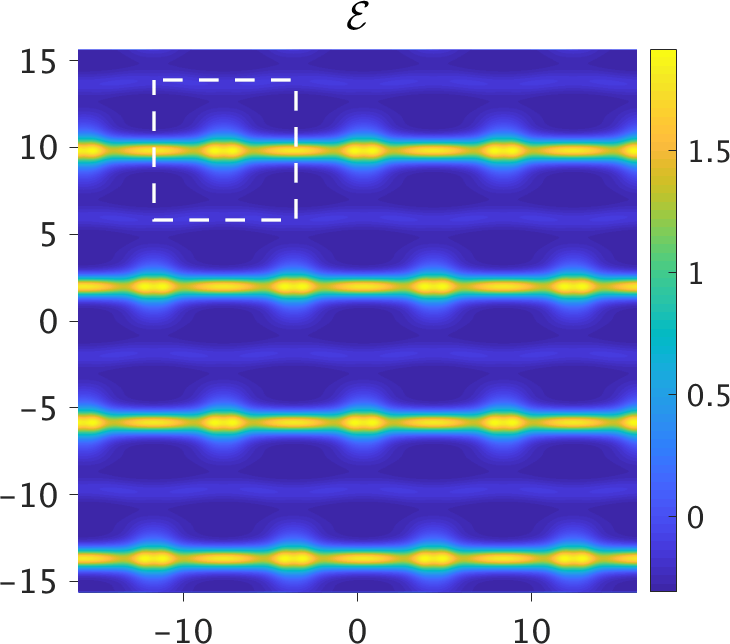}\hfill
  \includegraphics[width=0.251\linewidth]{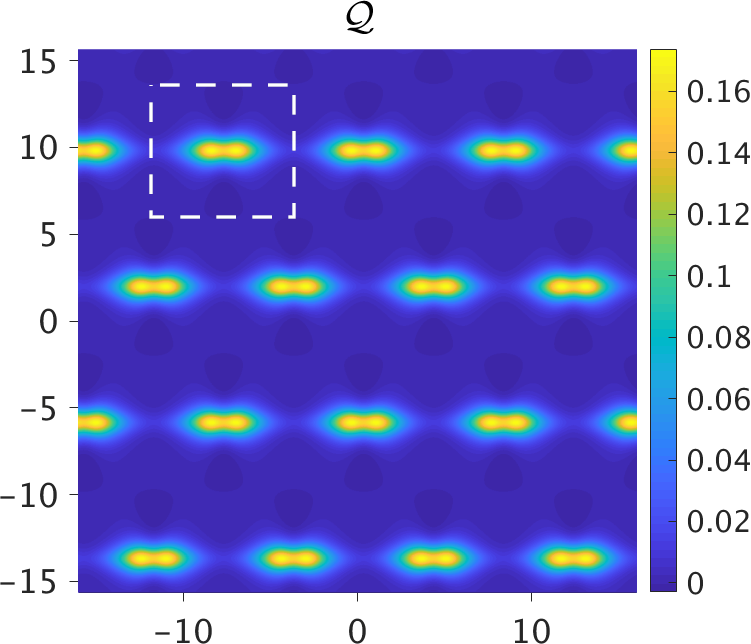}
  \put(-510,108){(b)}\\
  \caption{CSL phase (weak condition) with excited bimerons on the
    DWs: (a) $K=0.5$, (b) $K=1$, both for $B=0.8$.
    The energies of the solutions in this figure are minimized with
    respect to the dimensions only in the $y$-direction subject to
    periodic boundary conditions.
    A single bimeron is marked with a dashed line in all panels.
    For details of the four columns of panels, see the caption of
    Fig.~\ref{fig:K0.42B0.8}.
  }
  \label{fig:K0.5_1B0.8}
\end{figure*}

\begin{figure*}
  \centering
  \includegraphics[width=0.218\linewidth]{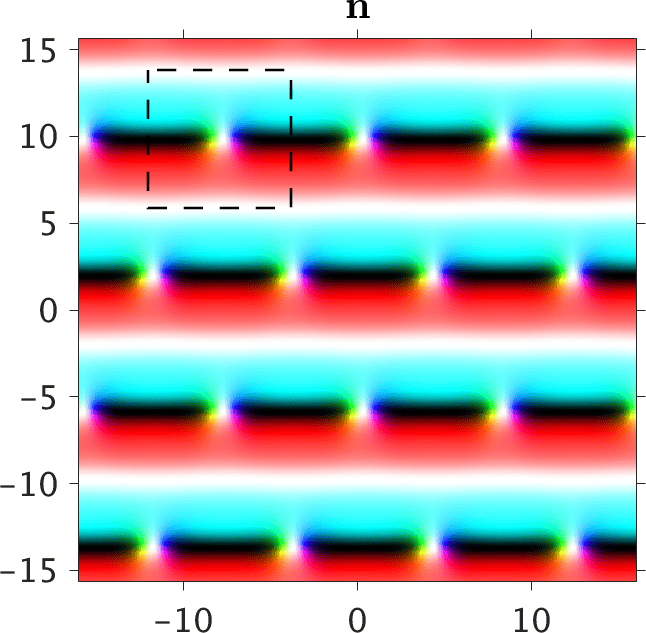}\hfill
  \includegraphics[width=0.252\linewidth]{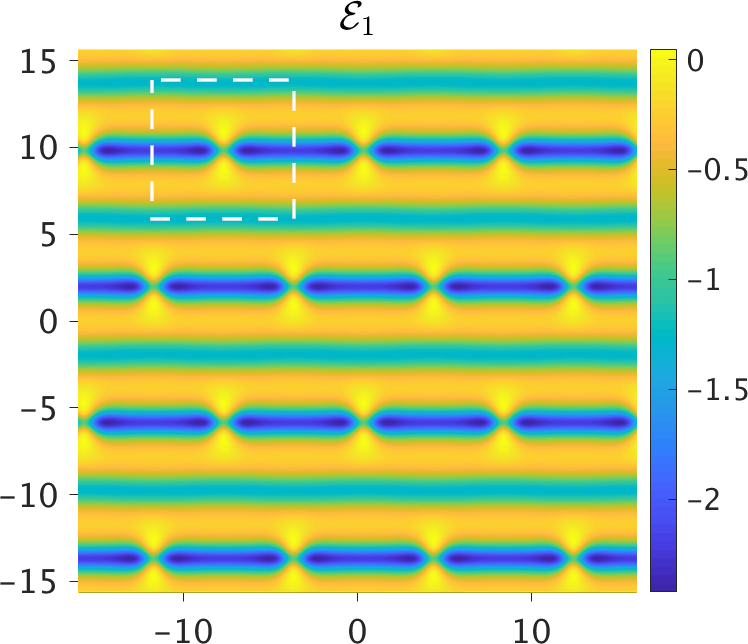}\hfill
  \includegraphics[width=0.238\linewidth]{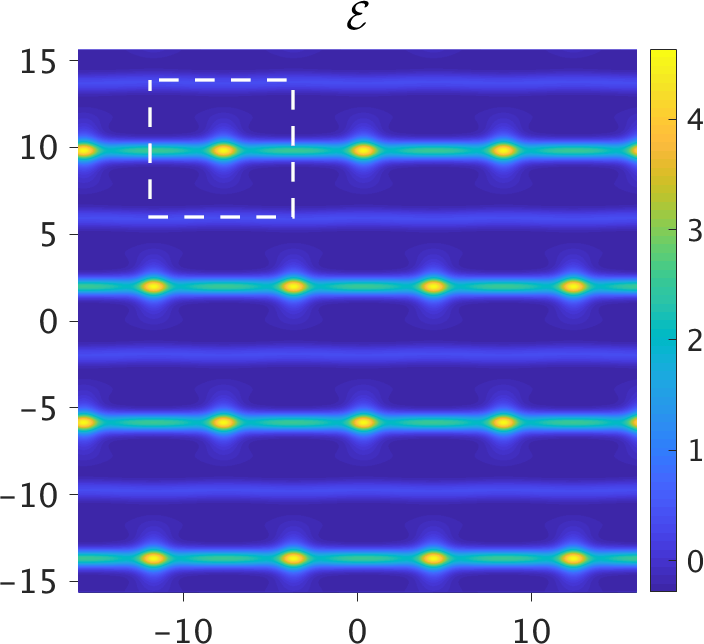}\hfill
  \includegraphics[width=0.252\linewidth]{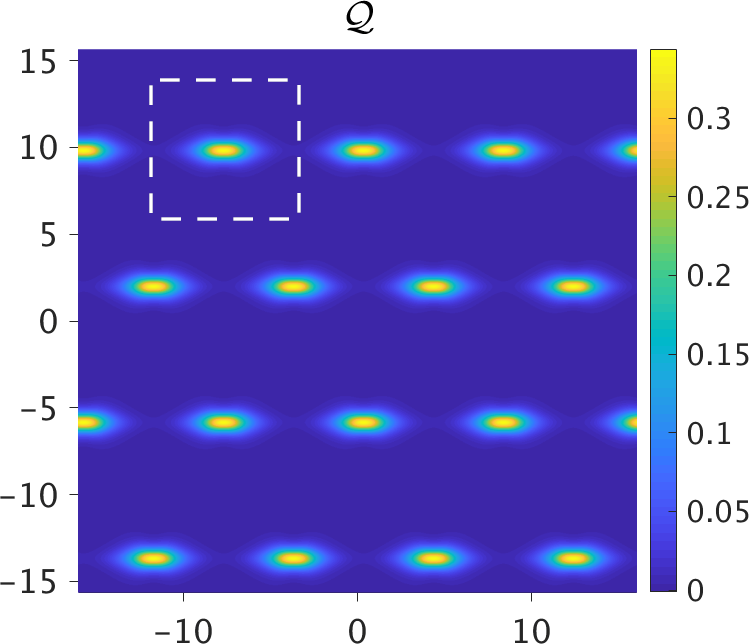}
  \put(-510,108){(a)}\\
  \includegraphics[width=0.218\linewidth]{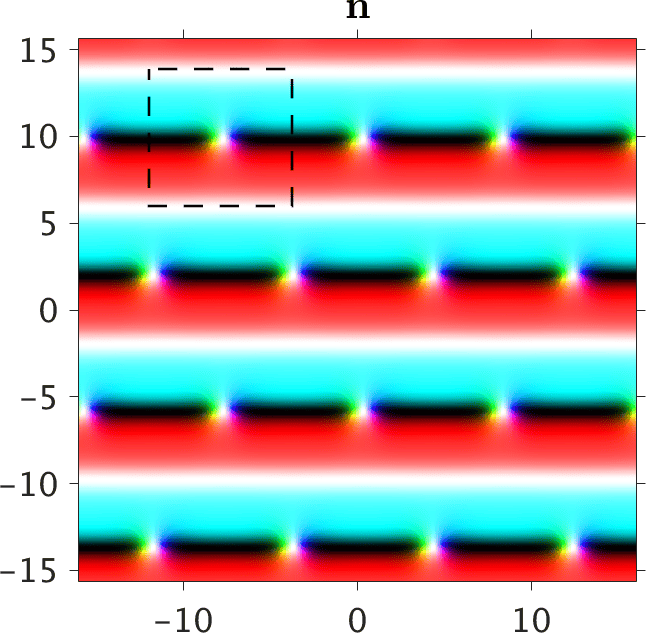}\hfill
  \includegraphics[width=0.252\linewidth]{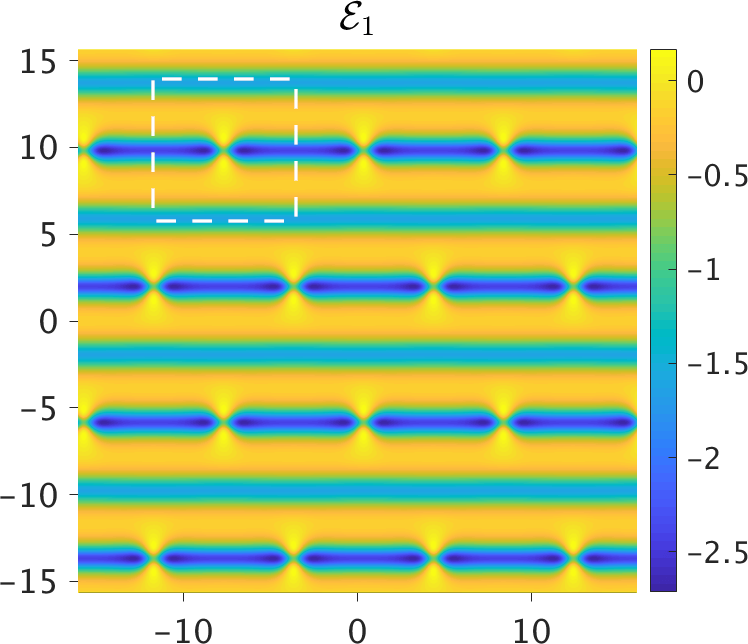}\hfill
  \includegraphics[width=0.237\linewidth]{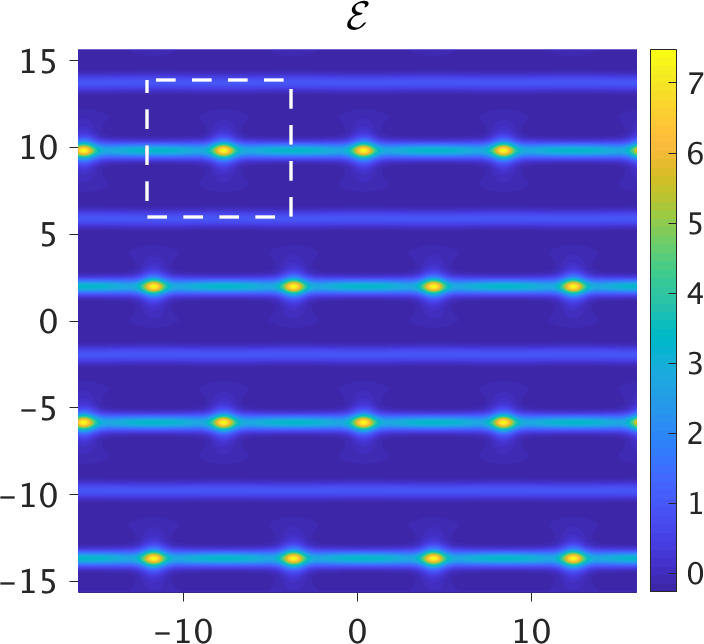}\hfill
  \includegraphics[width=0.246\linewidth]{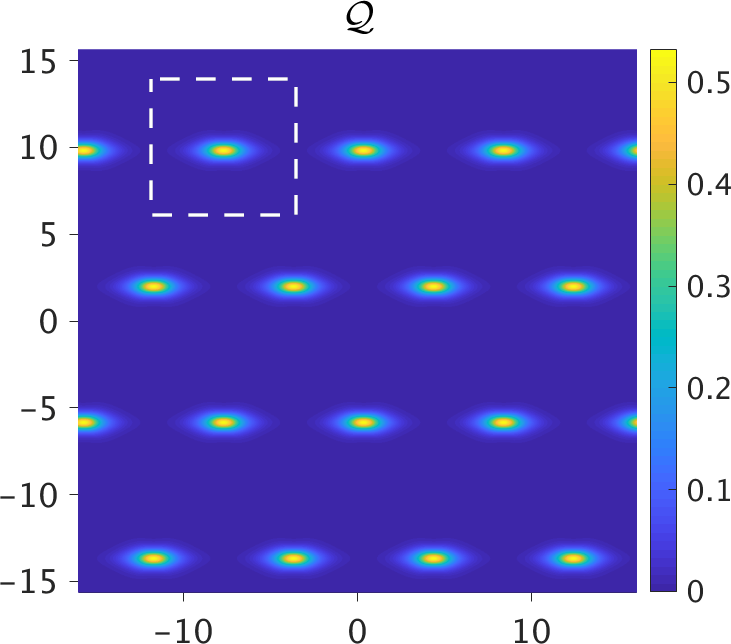}
  \put(-510,108){(b)}\\
  \caption{(Excited) DWs in the ferromagnetic phase hosting excited
    bimerons: (a) $K=1.5$, (b) $K=2$, both for $B=0.8$.
    A single bimeron is marked with a dashed line in all panels.
    For details of the four columns of panels, see the caption of
    Fig.~\ref{fig:K0.42B0.8}.
  }
  \label{fig:K1.5_2B0.8}
\end{figure*}

If instead of decreasing the anisotropy coefficient, $K$, we increase
it from the CSL-on-CSL point, i.e.~from $K=0.42$ to $K=0.5$, the
bimeron is no longer the ground state on the DW, but an excited state,
see Fig.~\ref{fig:K0.5_1B0.8}(a) [the bimeron is again marked with a
  dashed line].
This means that the number of bimerons placed the DW is now not
related to the length of the DW, and hence the more bimerons (per unit
length) the higher the energy.
Therefore, we keep the number of bimerons fixed when increasing the
anisotropy coefficient $K$: this allows us to illustrate the size of
the gaps between the DWs.
As an illustration we show two cases of the weak condition's CSL with
excited bimerons at two different values of $K=0.5$ and $K=1$ in
Fig.~\ref{fig:K0.5_1B0.8}(a) and (b), respectively.
In the former figure (panel (a)) the DWs are short and look like very
deformed skyrmions, whereas in the latter (panel (b)) the bimerons are
now visible as a skyrmionic soliton in the field $\bn$ (see the
left-most panel of the figure) and the topological charge is located
around the ``rupture'' of the DW, but is still visibly 2 separate
topological 1/2-skyrmion charges.

Increasing the value of the anisotropy coefficient $K$ further, we
leave the CSL (with weak condition) and enter the ferromagnetic DW
phase, where the DW is now an excited state with respect to the
homogeneous vacuum state, see Fig.~\ref{fig:phasediagram}(a).
For ease of comparison of the figures, we keep the same DW-anti-DW
separation distance (in the $y$-direction), but this is now not
determined by energy minimization as it was for smaller $K$ (indeed,
diluting the system by separating the DW and the anti-DW will lower
the energy).
Illustrating what happens to the bimeron, that continues to be an
excited state on the DW, we show in Fig.~\ref{fig:K1.5_2B0.8}(a) and
(b) the bimerons for $K=1.5$ and $K=2$, respectively [the bimeron is
  marked with a dashed line].
The bimeron is now confined in both cases, see the right-most panel of
the figures -- that is, the two 1/2-skyrmion topological charge
densities are no longer visibly separated along the DW, but the
topological charge looks now similar to a deformed skyrmion's charge.


\begin{figure*}[!htp]
  \centering
  \includegraphics[width=0.224\linewidth]{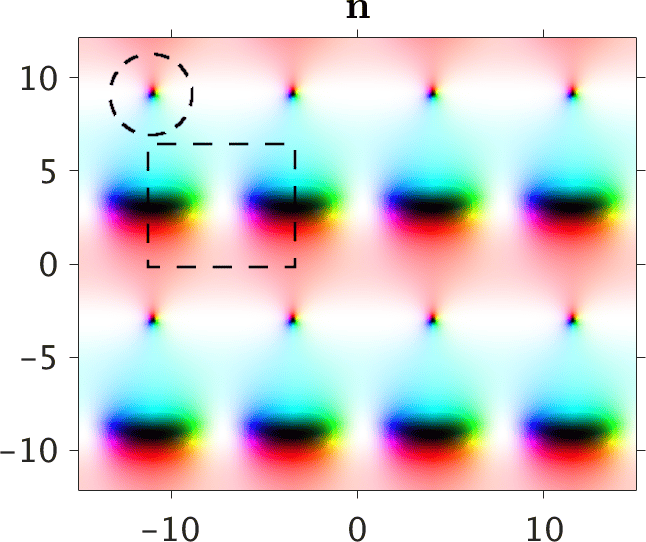}\hfill
  \includegraphics[width=0.25\linewidth]{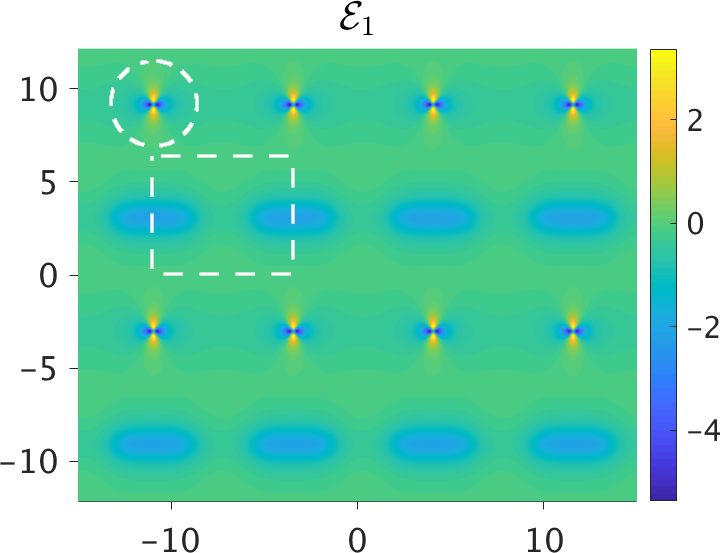}\hfill
  \includegraphics[width=0.251\linewidth]{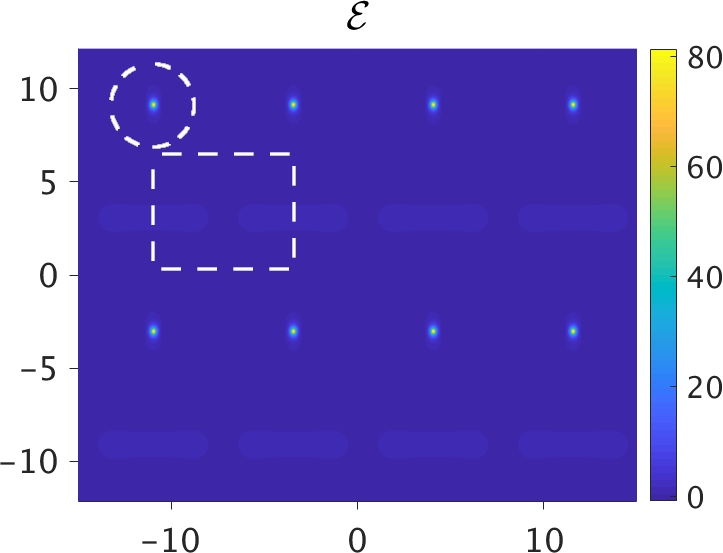}\hfill
  \includegraphics[width=0.25\linewidth]{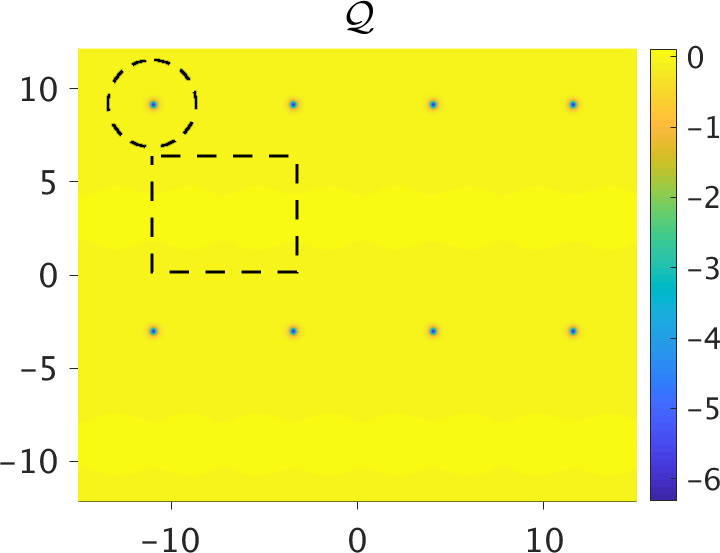}
  \caption{CSL on CSL phase (weak condition) with excited
    anti-skyrmion on the anti-DWs: $K=0.73$, $B=1.05$.
    A single bimeron is marked with a dashed square and a single
    anti-skyrmion is marked with a dashed circle, in all panels.
    For details of the four columns of panels, see the caption of
    Fig.~\ref{fig:K0.42B0.8}.
  }
  \label{fig:K0.73B1.05_anti-skyrmion}
\end{figure*}

\begin{figure*}[!htp]
  \centering
  \includegraphics[width=0.218\linewidth]{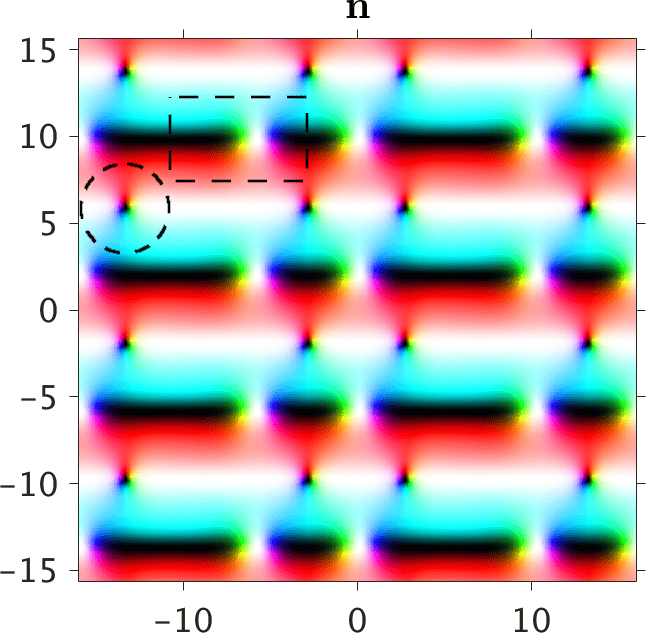}\hfill
  \includegraphics[width=0.244\linewidth]{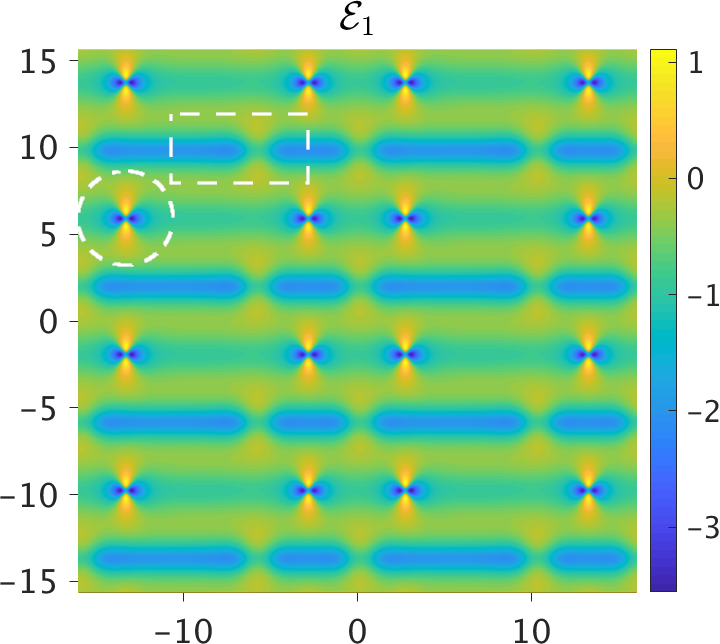}\hfill
  \includegraphics[width=0.244\linewidth]{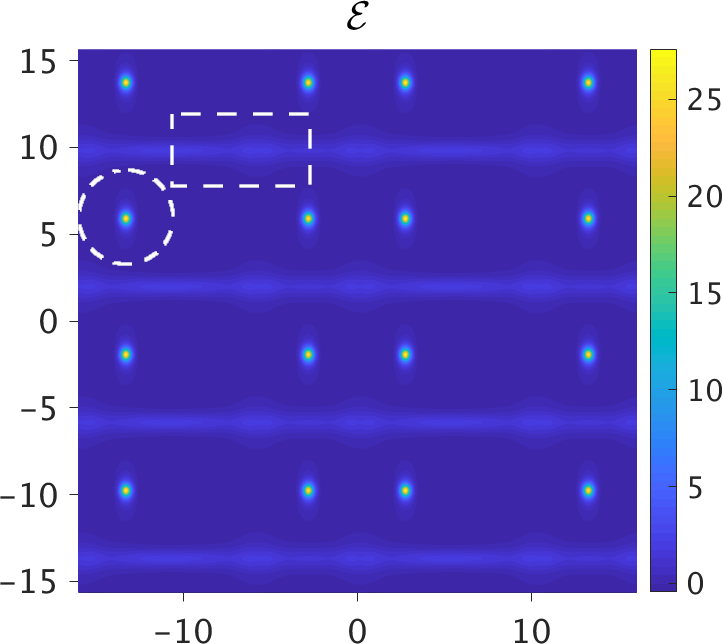}\hfill
  \includegraphics[width=0.252\linewidth]{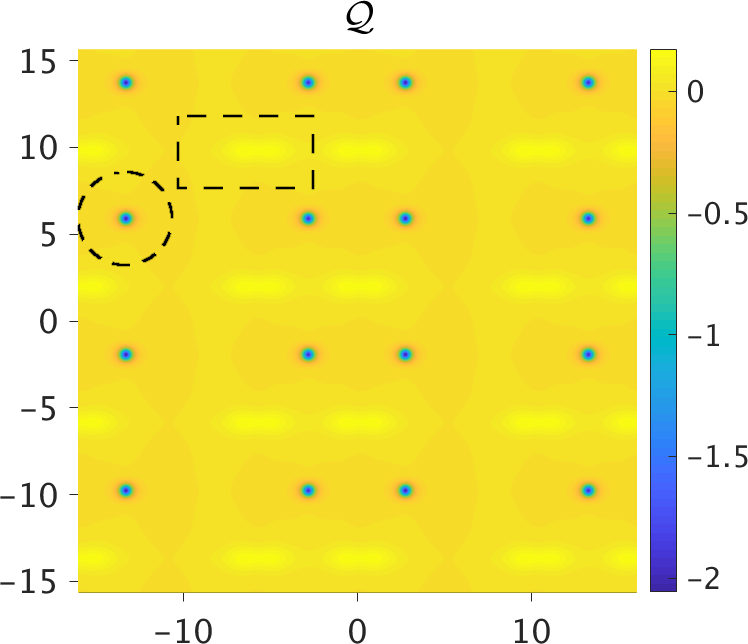}
  \caption{Ferromagnetic phase with bimerons on the DWs and anti-skyrmions
    on the anti-DWs: $K=1$, $B=0.8$.
    A single bimeron is marked with a dashed square and a single
    anti-skyrmion is marked with a dashed circle, in all panels.
    For details of the four columns of panels, see the caption of
    Fig.~\ref{fig:K0.42B0.8}.
  }
  \label{fig:K1B0.8_anti-skyrmion}
\end{figure*}

We now turn to the phase anti-skyrmion, whose existence is shown in
the phase diagram \ref{fig:phasediagram}(b) with a dark-green overlay
on the DW's soliton phase diagram.
The anti-skyrmion, as mentioned above, exists only on the anti-DW
which is always of type $\theta_S^-$ (see Tab.~\ref{tab:Q}), it has a
\emph{negative} topological charge, $Q=-|Q_{\rm sh}|$ (as opposed to the
bimeron that has a positive topological charge, $Q=|Q_{\rm sh}|$),
it exists only for ``large'' anisotropy, it is much smaller than the
bimeron and finally it is not deformed much by the anisotropic
potential term, see Figs.~\ref{fig:K0.73B1.05_anti-skyrmion} and
\ref{fig:K1B0.8_anti-skyrmion}.
In Fig.~\ref{fig:K0.73B1.05_anti-skyrmion} we illustrate the
anti-skyrmion on the anti-DW as an excited state of the CSL on a CSL
with the weak condition that we previously presented in
Fig.~\ref{fig:K0.73_0.65B1.05}(a).
We have marked the anti-skyrmion with a circular dashed line and as
always the bimeron is marked with a rectangular dashed line in all
panels of the figure.
The anti-skyrmion feels the presence of the bimeron and indeed it
prefers to place itself near one of the two merons.
In the case shown in Fig.~\ref{fig:K0.73B1.05_anti-skyrmion}, the DW
is so short that the anti-skyrmion is aligned roughly by the middle of
the DW.
Due to symmetry (i.e.~$y\to-y$), the lattice becomes quadratic in
contradistinction to the (stretched) hexagonal lattice of bimerons in
the same point of parameter space, but without the anti-skyrmion, shown in
Fig.~\ref{fig:K0.73_0.65B1.05}(a).

Another example of the anti-skyrmion in co-existence with the bimeron
is shown in Fig.~\ref{fig:K1B0.8_anti-skyrmion}, which is the same
point in parameter space as presented previously in
Fig.~\ref{fig:K0.5_1B0.8}(b), albeit without the anti-skyrmion.
The anti-skyrmion is again marked with a circular dashed line and the
bimeron is marked with a rectangular dashed line in all panels of the
figure.
We are now in the ferromagnetic phase, so the both the DW-anti-DW
system is an excited state and the bimerons, as well as the
anti-skyrmions, are excited states.
Therefore, we do no minimize the energy with respect to the density of
any of the solitons ($x$- or $y$-direction).
Without the presence of the anti-skyrmion, the bimeron lattice
organizes itself in a hexagonal pattern, see
Fig.~\ref{fig:K0.5_1B0.8}(b).
However, the anti-skyrmion is attracted to one of the merons, which
aligns the meron and the anti-skyrmion in the $y$-direction, but this
also creates a tension for the DWs.
Indeed, the DWs break the periodic symmetry in the $x$-direction by
making every second DW longer.
This way, 2 out of 4 anti-skyrmions are in the vicinity of two merons
and not just a single meron, see Fig.~\ref{fig:K1B0.8_anti-skyrmion}.


\begin{figure*}[!htp]
  \centering
  \includegraphics[width=0.218\linewidth]{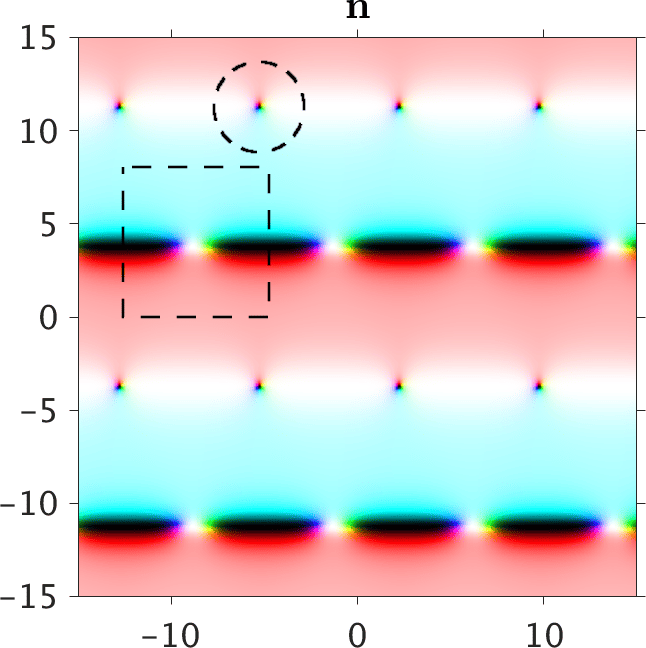}\hfill
  \includegraphics[width=0.243\linewidth]{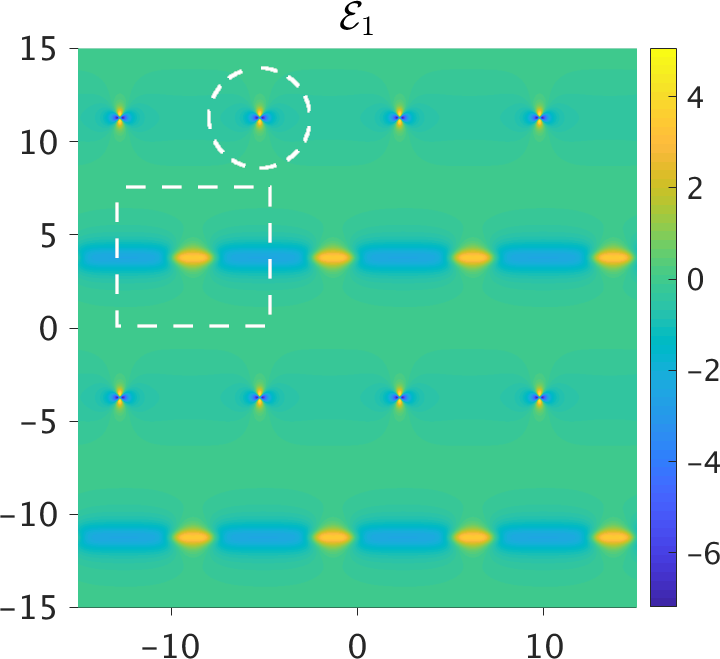}\hfill
  \includegraphics[width=0.251\linewidth]{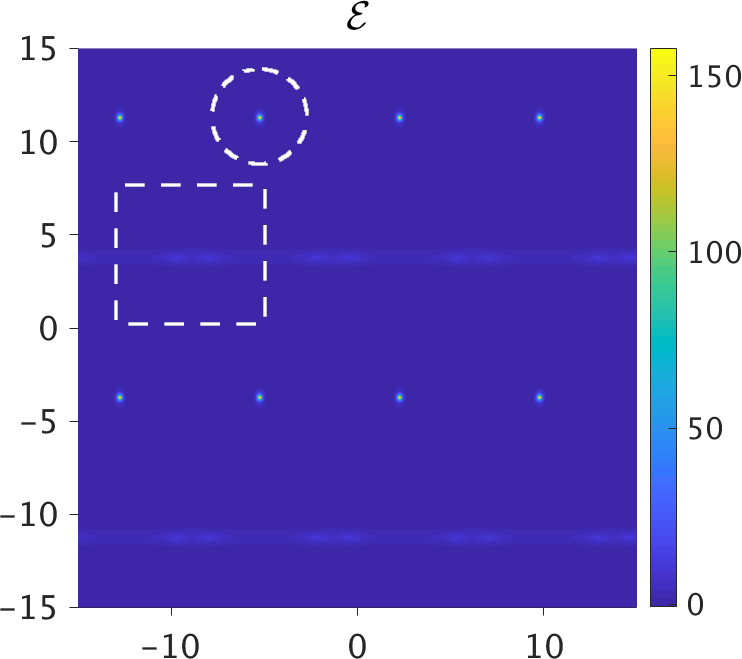}\hfill
  \includegraphics[width=0.249\linewidth]{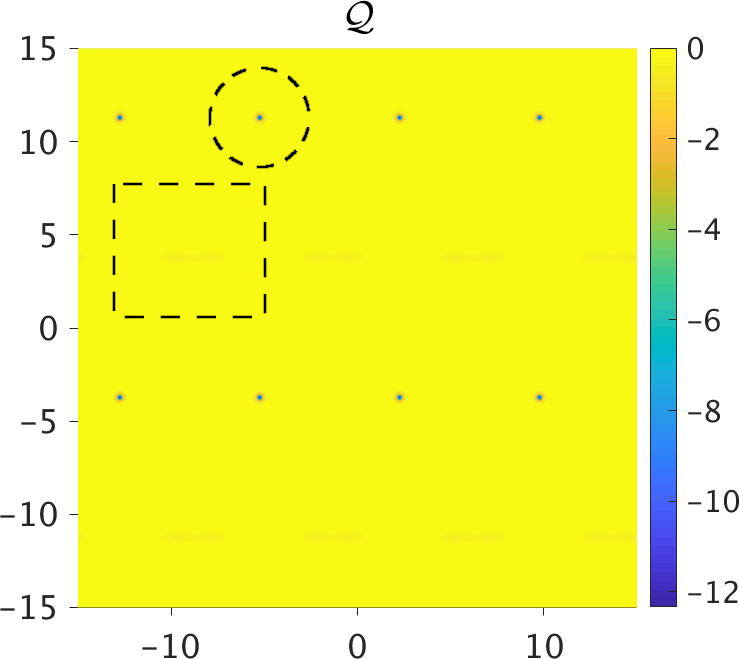}
  \caption{Ferromagnetic phase with anti-bimerons on the DWs and
    anti-skyrmions on the anti-DWs: $K=1$, $B=1.41$.
    A single anti-bimeron is marked with a dashed square and a single
    anti-skyrmion is marked with a dashed circle, in all panels.
    For details of the four columns of panels, see the caption of
    Fig.~\ref{fig:K0.42B0.8}.
  }
  \label{fig:K1B1.41_anti-bimeron_anti-skyrmion}
\end{figure*}

The next excited state that we will discuss is the anti-bimeron, which
is shown with a gray-colored region in the phase diagram
\ref{fig:phasediagram}(a), which resides at $K=1$ and $B=1.41$, 
see Fig.~\ref{fig:K1B1.41_anti-bimeron_anti-skyrmion}.
The anti-skyrmion is again marked with a circular dashed line and the
\emph{anti-}bimeron is now marked with a rectangular dashed line in
all panels of the figure.
The \emph{anti-}bimeron is conceptually identical to the bimeron,
except that the magnetization vector covers the colored spectrum of
the bimeron \emph{counter}-clockwise as opposed to in a clockwise
orientation.
The two solitons are related by a parity transformation, but as we
know, parity is broken by the DMI.
Bimeron or anti-bimeron, the anti-skyrmion is still attracted.
Therefore, the anti-skyrmion on the anti-DW is aligned vertically with
one of the merons on the DW: since the DWs are quite short in the
example of Fig.~\ref{fig:K1B1.41_anti-bimeron_anti-skyrmion}, the
anti-skyrmion is horizontally roughly placed near the middle of the
DW.

\begin{figure*}[!htp]
  \centering
  \includegraphics[width=0.35\linewidth]{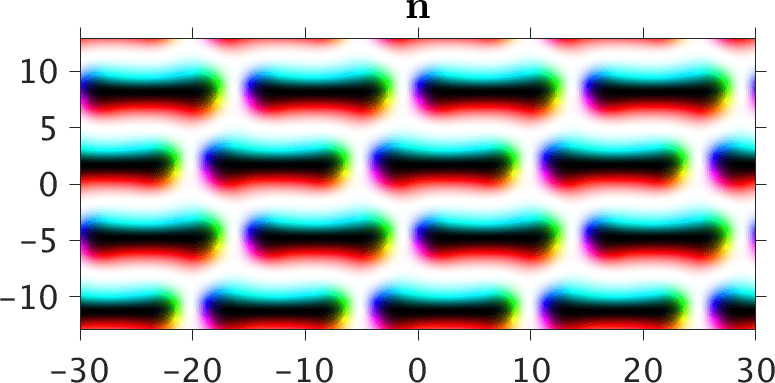}\ \ \qquad\qquad
  \includegraphics[width=0.4\linewidth]{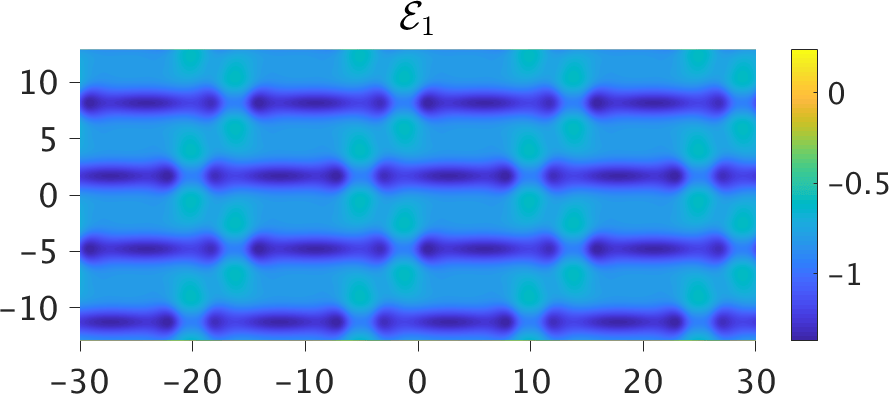}\\[5pt]
  \includegraphics[width=0.4\linewidth]{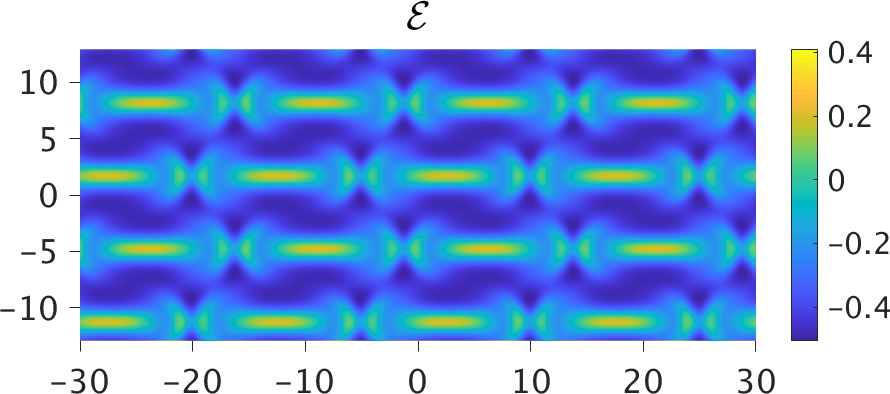}\qquad
  \includegraphics[width=0.4\linewidth]{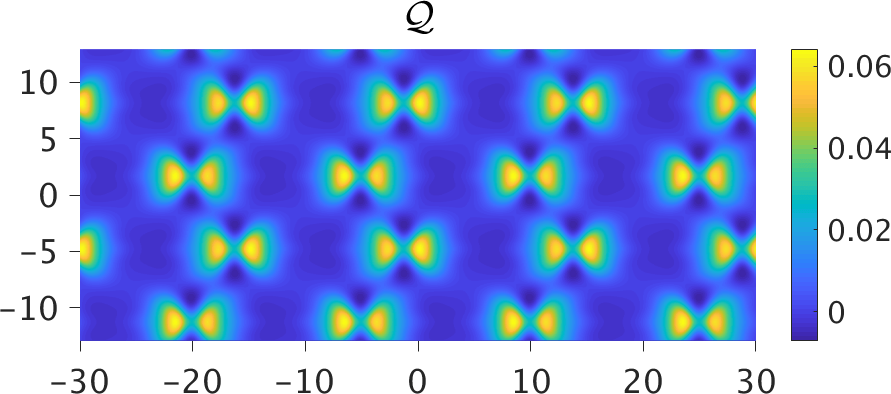}
  \caption{Bimeron crystal phase: $K=0.14$, $B=0.3$.
    For details of the four panels, see the caption of
    Fig.~\ref{fig:K0.42B0.8}.
  }
  \label{fig:K0.14B0.3_bimeron}
\end{figure*}

Finally, we move back into the ``no DW'' part of the phase diagram and instead of looking at the known skyrmion crystal (SkX) phase, we move downward in the phase diagram to the bimeron crystal phase, in particular to the point $K=0.14$, $B=0.3$, see Fig.~\ref{fig:phasediagram}(b). 
Technically, there exists no DW wall in this point of the phase diagram since $K<B/2$. Nevertheless, the solutions look like bimerons on DWs, see Fig.~\ref{fig:K0.14B0.3_bimeron}.
We call this deformation of the skyrmion crystal the bimeron crystal phase. That is, the DW is not mathematically possible, but the deformation (stretching) of the skyrmion in the skyrmion crystal is possible. 
What distinguishes the bimeron crystal phase from the squashed skyrmions in the SkX phase is that in the bimeron crystal phase the skyrmions are concave or in more geometrical terms: the curvature is not of the same sign all the way around the blob of anti-vacuum.

\begin{figure*}[!htp]
  \centering
  \includegraphics[width=0.35\linewidth]{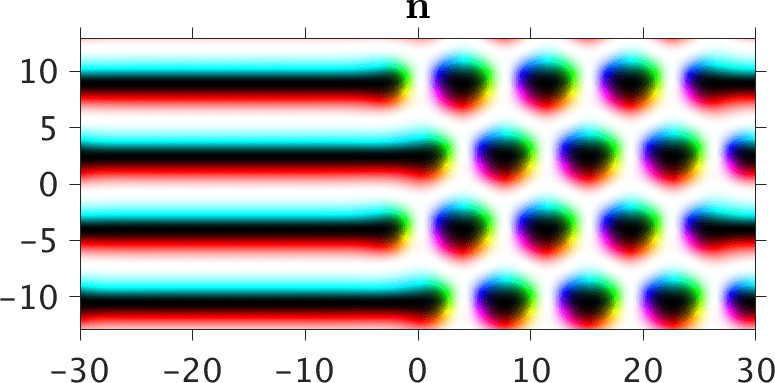}\ \ \qquad\qquad
  \includegraphics[width=0.4\linewidth]{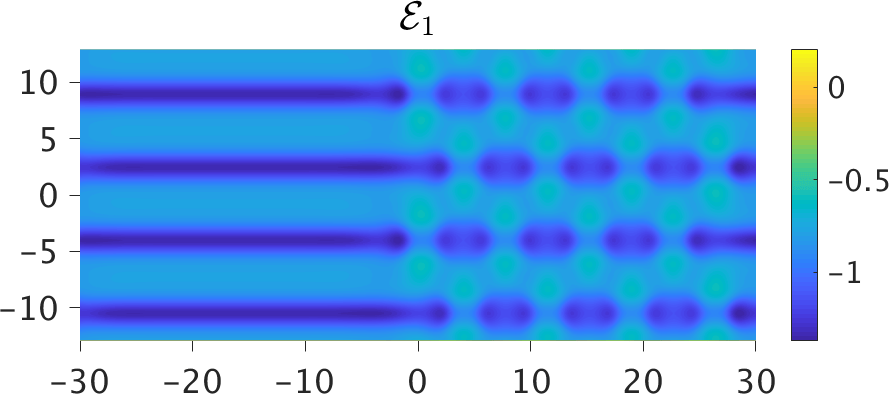}\\[5pt]
  \includegraphics[width=0.4\linewidth]{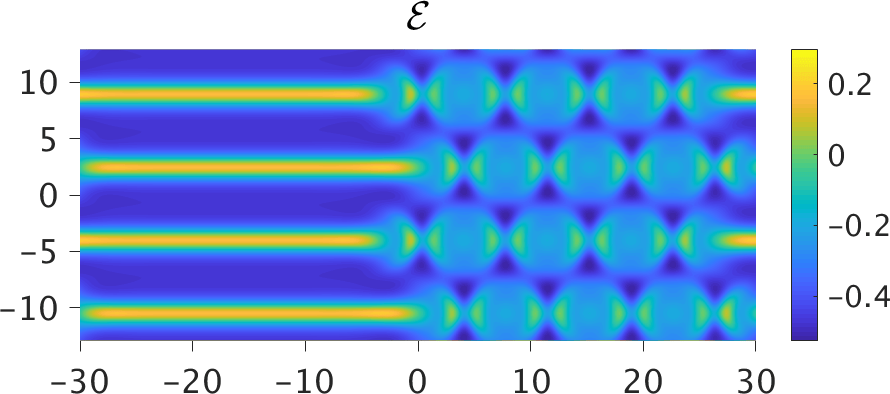}\qquad
  \includegraphics[width=0.4\linewidth]{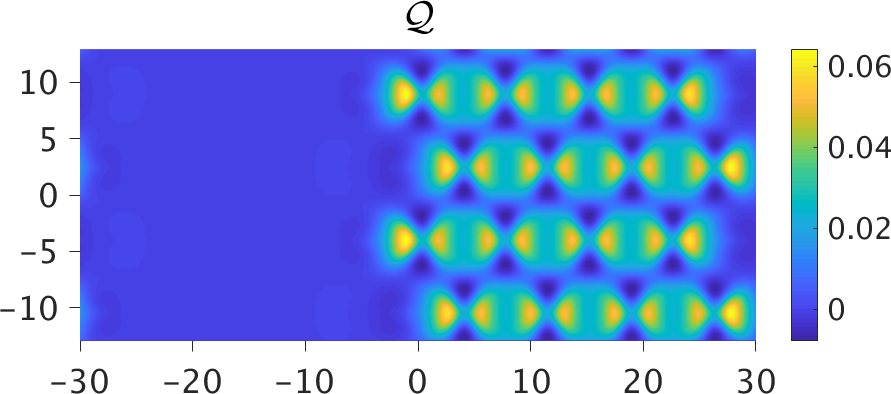}
  \caption{Bimeron and skyrmion crystal mixed phase: $K=0.14$, $B=0.3$.
    For details of the four panels, see the caption of
    Fig.~\ref{fig:K0.42B0.8}.
  }
  \label{fig:K0.14B0.3_bimeron_skyrmion_mixed}
\end{figure*}

Interestingly, the crystal of bimerons is not the only solution at this point in the phase diagram, but a mixed-phase solution exists with almost the same energy density as the bimeron crystal, but actually slightly lower energy, and it is a very stretched skyrmion that resembles a DW very much (even though DWs are not mathematically existing in this part of the phase diagram) coexisting with a skyrmion crystal, see Fig.~\ref{fig:K0.14B0.3_bimeron_skyrmion_mixed}.
Interestingly, the skyrmion crystal with round skyrmions does not exist in this part of the phase diagram due to the large anisotropy coefficient $K$, but it can coexist with a bimeron (DW look-alike) or rather a very stretched skyrmion that makes up the constituent of the bimeron crystal phase.
There is no energetical minimum of this solution as the bimeron (stretched skyrmion) can continue to stretch to minimize the total energy density.
However, the round skyrmion is not locally preferred and the DW is not existing this part of the phase diagram, so this mixed phase is somehow fooling the rules of the game.

\begin{figure*}[!htp]
  \centering
  \includegraphics[width=0.355\linewidth]{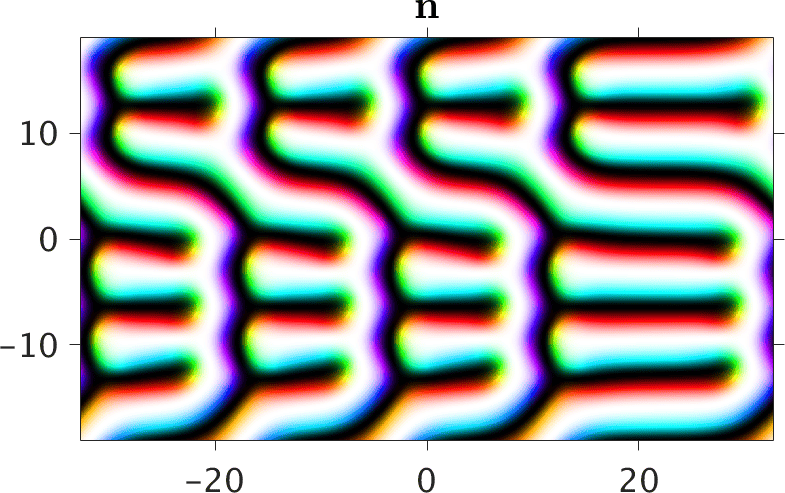}\ \qquad\qquad
  \includegraphics[width=0.40\linewidth]{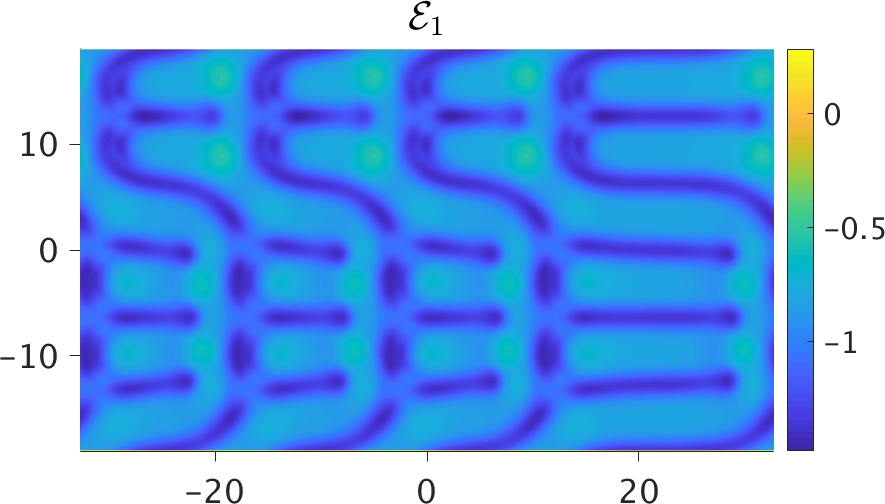}\ \ \ \\[5pt]
  \includegraphics[width=0.40\linewidth]{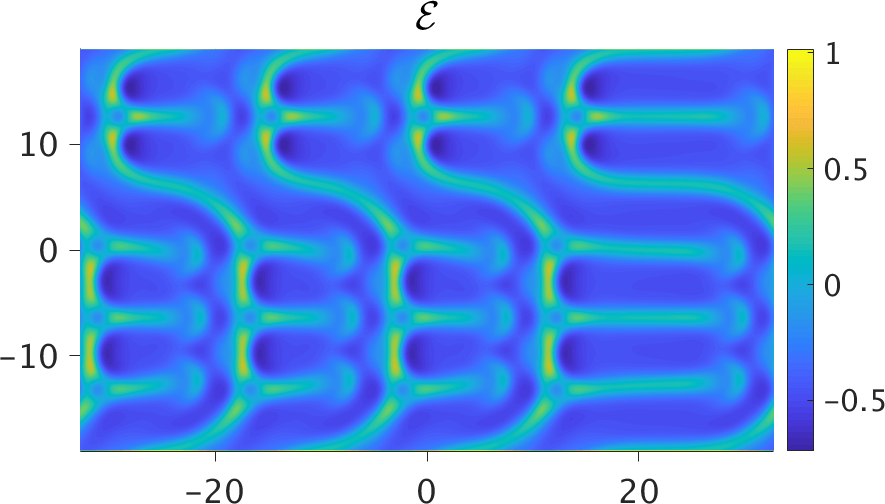}\qquad
  \includegraphics[width=0.407\linewidth]{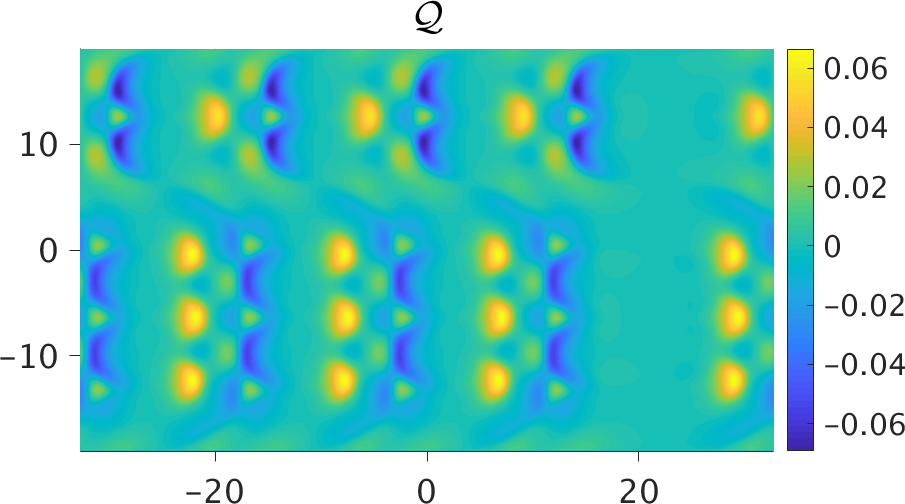}
  \caption{``Chiral spiral-type'' excited state: $K=0.14$, $B=0.3$.
    For details of the four panels, see the caption of
    Fig.~\ref{fig:K0.42B0.8}.
  }
  \label{fig:K0.14B0.3_chiral_spiral}
\end{figure*}

Finally, we should mention that there is a plethora of ``chiral spiral-type'' non-topological solutions ($Q=0$), and an example at the same point in the phase diagram, $K=0.14$, $B=0.3$, there is a non-topological solution with very complicated local structure, see Fig.~\ref{fig:K0.14B0.3_chiral_spiral}.
There are many kinds of these non-topological solutions in the skyrmion crystal (SkX) phase, which we, however, have not mapped out systematically.
A major reason for this is that it is hard to find a suitable Ansatz for their initial condition and it is not clear which kind has the lowest energy.
Nevertheless, we find that all these non-topological solutions, including that of Fig.~\ref{fig:K0.14B0.3_chiral_spiral}, are all excited states compared with the skyrmion crystal or the bimeron crystal phase, although without a systematic exploration, we cannot state this with certainty.

Let us finally mention a curious feature of the phase diagram visible in Fig.~\ref{fig:phasediagram}(b), which is that the bands of stable bimerons in the DW-phase (i.e.~for $K>B/2$) continue into the ``no-DW'' phase (i.e.~$K<B/2$). 
The upper boundary of existence of the bimeron in the DW-phase incarnates itself as the boundary between the skyrmion crystal (SkX) and the ferromagnetic skyrmion phase in the ``no-DW'' phase, where skyrmions are excited states, whereas the lower boundary of existence of the bimerons in the DW-phase turns into the boundary between the SkX and the bimeron crystal phase in the ``no-DW'' phase.
This further strengthes the claim that the transition between the DW-phase and the ``no-DW'' phase is really just a crossover and not a phase transition. 

\section{Conclusion and discussion}\label{sec:conclusion}

In this paper, we have studied the chiral magnetic system with a Bloch
DMI, corresponding to the Dresselhaus SOC, with an in-plane anisotropy
and an out-of-plane Zeeman term.
We exhaust the possibilities of placing kinks or anti-kinks on DWs and
anti-DWs in a large parameter space $K\in[0,2]$ by $B\in[0,2]$ (with
the point $K=B=0$ excluded) and discover for the first time a CSL on a
CSL, an anti-bimeron as an excited state on the DW, and finally
excited states with coexisting anti-skyrmions and (anti-)bimerons.
We find that the CSL on the CSL phase is not separated from the
skyrmion crystal lattice phase (SkX), but is smoothly connected via a
crossover.
We further observed that in the CSL on a CSL or the DW with a CSL (kink
lattice) phases, the bimerons that live on the DWs feel each other's
presence even though the DWs are separated by an anti-DW, which
results in a hexagonal lattice of bimerons, that however is stretched
in the vertical direction (i.e.~the direction orthogonal to the DW).
On the other hand, we observed that the anti-skyrmions tend to align
themselves with at least one meron or the entire bimeron -- if the DW
is short enough.
This creates a vertical alignment and often quadratic lattices, as
opposed to the usual hexagonal lattices observed in skyrmionic
systems.

Our study has been exhaustive within the potential parameters and by
rescaling it includes any choice of the DMI coupling as well as any
choice of Heisenberg interchange interaction.
However, we have not considered the effect due to demagnetization that
could potentially be important for small magnetic field in case of
N\'eel-type DMI, whereas they usually are not so important for
Bloch-type DMI \cite{Gudnason:2025gqs}, viz.~the DW and the
isolated skyrmion does not change size due to demagnetization effects
in Bloch-type DMI systems.
The main reason for not taking the demagnetization effect into account
in this work, where we perform an exhaustive scan of the parameter
space is that the parameter space is only 2-dimensional without the
demagnetization field.
Since the coupling to the demagnetization field is fixed by the
magnetic permeability, we can no longer scale away the DMI coupling --
this makes the parameter space 3-dimensional when taking into account
demagnetization!
We can say that for small demagnetization or larger DMI couplings or
for Bloch-type systems, our results hold -- but in other cases, some
minor changes may occur.
Generally speaking, however, the demagnetization field often only
changes the sizes of the static ground state, but not qualitatively
the properties of the state.

It would also be very interesting to consider finite temperature, but
this also adds a dimension -- cranking up the parameter space to a
4-dimensional one. Needless to say, this requires much more work in
order to perform an exhaustive study of the possible phases.
We thus leave these interesting, albeit computationally expensive
extensions for future work.

\subsection*{Acknowledgments}

We thank Yuki Amari for useful comments 
and collaboration at the early stage of this work.
S.~B.~G.~thanks the Outstanding Talent Program of Henan University for
partial support. 
This work is supported in part by JSPS KAKENHI [Grants No.~JP22H01221 (M.~N.), 
  JP23K22492 (M.~N.)] and the WPI program
``Sustainability with Knotted Chiral Meta Matter (WPI-SKCM$^2$)'' at
Hiroshima University (M.~N.).

\bibliography{references}

\end{document}